\documentclass[journal=jacsat,manuscript=article]{achemso}
\setkeys{acs}{articletitle=true}
\SectionNumbersOn
\usepackage[T1]{fontenc} 
\usepackage[hidelinks]{hyperref}
\usepackage{easyReview}
\mciteErrorOnUnknownfalse
\DeclareUnicodeCharacter{2212}{-}
\usepackage[normalem]{ulem}
\usepackage{xcolor}
\usepackage{amsmath}
\author{Sakthi Priya Amirtharaj}
\author{Zhiyuan Xie}
\author{Josephine Si Yu See}
\author{Gabriele Rolleri}
\author{Konstantin Malchow}
\author{Wen Chen}
\affiliation
{Institute of Physics, Ecole Polytechnique Fédérale de Lausanne (EPFL), CH-1015 Lausanne, Switzerland}
\author{Alexandre Bouhelier}
\affiliation
{Laboratoire Interdisciplinaire Carnot de Bourgogne CNRS UMR 6303, Université de Bourgogne, 21000 Dijon, France}
\author{Emanuel Lörtscher}
\affiliation
{IBM Research Europe - Zurich, Säumerstrasse 4, CH-8803 Rüschlikon , Switzerland}
\author{Christophe Galland}
\affiliation
{Institute of Physics, Ecole Polytechnique Fédérale de Lausanne (EPFL), CH-1015 Lausanne, Switzerland}
\email{chris.galland@epfl.ch}

\title[]
{Light Emission and Conductance Fluctuations in Electrically Driven and Plasmonically Enhanced Molecular Junctions}

\begin{document}
\renewcommand{\baselinestretch}{1.3}\small
\begin{abstract}
\textbf{
Electrically connected and plasmonically enhanced molecular junctions combine the optical functionalities of high field confinement and enhancement (cavity function), and of high radiative efficiency (antenna function) with the electrical functionalities of molecular transport. \add{Such combined optical and electrical probes have proven useful for the fundamental understanding of metal-molecule contacts} and contribute to the development of nanoscale optoelectronic devices including ultrafast electronics and nanosensors. 
Here, we \add{employ a self-assembled metal-molecule-metal junction with a nanoparticle bridge to }investigate correlated fluctuations in conductance and tunneling-induced light emission at room temperature. \add{Despite the presence of hundreds of molecules in the junction,} the electrical conductance and light emission are both highly sensitive to atomic-scale fluctuations -- a phenomenology reminiscent of picocavities observed in Raman scattering and of luminescence blinking from photo-excited plasmonic junctions. \add{Discrete steps in conductance associated with fluctuating emission intensities through the multiple plasmonic modes of the junction are consistent with a finite number of randomly localized, point-like sources dominating the optoelectronic response. }Contrasting with these microscopic fluctuations, the overall plasmonic and electronic functionalities of our devices feature long-term survival at room temperature and under an electrical bias of a few volts, allowing for measurements over several months.}
\end{abstract}
\vspace{2mm}

\section*{Introduction}
Plasmonic nanocavities with extreme light confinement allow electromagnetic interactions with few or even single molecules to be studied and tailored \cite{maccaferri_recent_2021, zhang_chemical_2013}. They can be electrically connected \cite{wang_molecular_2016}, resulting in molecular junctions that provide opportunities to measure both molecular transport \cite{xiang_molecular-scale_2016,  gehring_single-molecule_2019, karthauser_integration_2020, kos_quantum_2021} and plasmonically enhanced optical signals. Such combined plasmonic and electrical experiments on molecular junctions are useful for a fundamental understanding of molecular properties and metal-molecule interfaces and for various applications including high-speed nanosized electronics \cite{wang_plasmonic_2022, aradhya_single-molecule_2013, zhou_ultrafast_2021}. For instance, electrical transport and spectroscopy of metal-molecule-metal junctions reveal quantum properties originating from structural and chemical rearrangements, observable even at ambient conditions \cite{gehring_single-molecule_2019}. 

While surface enhanced Raman scattering (SERS) is the most popular spectroscopic technique employed on molecular junctions \cite{ward_electromigrated_2007,natelson_nanogap_2013,suzuki_effect_2016, iwane_molecular_2017, liao_quantifying_2023},  electrically driven light emission is a more direct probe of the metal-molecule interface involved in both molecular transport and plasmon-enhanced light emission \cite{parzefall_optical_2019, downes_measurement_2002,buret_spontaneous_2015,cui_electrically_2020,roslawska_atomic-scale_2021, deeb_electrically_2023}. \add{ This broadband emission resulting from inelastic electron scattering can be understood as originating from the optical-frequency quantum shot noise of electrons tunneling across the potential barrier formed by the molecular junction \cite{lambe_light_1976, hone_theory_1978, schneider_optical_2010, fevrier_tunneling_2018}.} Such inelastic electron tunneling (IET) emission can efficiently couple to the plasmonic modes of the host nanocavity \cite{kern_electrically_2015}. The electrical and optical signals of these nanoscale devices contain information about the metal-molecule contacts as demonstrated by tip-enhanced Raman spectroscopy (TERS) and scanning-tunneling microscopy (STM) experiments, including STM-induced luminescence \cite{schull_electron-plasmon_2009, schneider_optical_2010,kuhnke_atomic-scale_2017,zhang_electrically_2017,liu_dramatic_2020, yang_sub-nanometre_2020,roslawska_atomic-scale_2021}. \add{Probing an invidual molecule usually} requires complex systems operating at ultra-high vacuum and/or cryogenic temperatures. 

\add{Here, we investigate single-molecule fluctuations within nanoscale plasmonic molecular junctions (PMJ) formed around a self-assembled monolayer (SAM) \cite{dadosh_measurement_2005,jafri_nano-fabrication_2015, karthauser_integration_2020,jeong_high-yield_2017}, which operate at room-temperature and feature long-term functionality. A SAM of thiol-terminated molecules is created on top of two lithographically defined electrodes bridged by a few nanoparticles. The idea of bridging nanoparticles between electrodes coated with SAM was proposed by Amlani et al. in 2002 \cite{amlani_approach_2002}. But for a long time, only the electrical properties of the molecular junctions were investigated. Kern\textit{ et al.} used a similar approach to trap nanoparticles with native ligands between pre-designed antenna structures to create a sub-nm gap and demonstrate electrically driven plasmonic emission by IET \cite{kern_electrically_2015}. The ligand molecules there merely played the role of an insulating spacer. 
Inspired by these approaches, we created electrodes with slanted sidewalls that are optimized for collection of the optical signal once a nanoparticle falls inbetween to form hybridized plasmonic resonances \cite{baumberg_extreme_2019}. A similar design was simulated, tested, and implemented as a dual-band nanoantenna in a previous work \cite{chen_continuous-wave_2021}.} These nanocavity junctions provide significant plasmonic enhancement that allows for fast optical measurements to probe atomic-scale fluctuations at the metal-molecule contact through its impact on light emission mediated by the tunneling of electrons across the gap. 
The junctions can be studied over weeks and months, allowing us to thoroughly investigate the impact of atomic fluctuations in the junction on both its photoemission and electrical transport characteristics. \add{ The devices continue to be dominated by molecular transport until they get damaged under too high voltage (several Volts), too large optical powers (mW/$\mu$m$^2$) or by electrostatic discharge (ESD).}

Our main finding is that a few randomly switching metal-molecule contact sites appear to control the macroscopic behavior of the device, despite the large number of molecules acting as a spacer. This microscopic dynamics is evidenced by discrete jumps in the light emission spectrum, as well as joint fluctuations of emission intensity and conductance that are consistent with a minimal model of fluctuating atom-molecule contacts. Similar point-like emission was recently evidenced in a large-area SAM tunnel junction between gold and eutectic gallium-indium alloy (EGaIn) contacts \cite{du_-chip_2016} where it was attributed to conformational changes in the molecule due to the excitation of vibrational modes, and no such blinking was observed in the current through the junction \cite{wang_operando_2019}. In Ref.\cite{du_-chip_2016} the resonance and polarization of the point-like plasmon source were modified by the applied bias voltage. With our junctions, we find no such shift in the plasmonic resonance with applied voltage. The plasmonic response is fixed by the nanoparticle-electrode cavity and is excited by electron tunneling through the junction, and we observe correlated blinking in both the light emission and conductance.  The spectral diffusion and intensity redistribution between the different plasmonic modes observed in our junctions resembles the movement of gold atoms locally modifying the emission efficiency of the individual plasmonic modes as demonstrated in photo-induced luminescence blinking of metal-molecule junctions \cite{chen_intrinsic_2021}. Our results therefore suggest that a phenomenology similar to that underlying the occurrence of picocavities in SERS \cite{benz_single-molecule_2016, carnegie_room-temperature_2018}, of blinking in gold nanojunction photoluminescence \cite{chen_intrinsic_2021} and of flickering in electronic Raman scattering \cite{carnegie_flickering_2020} can also be driven by electrical bias. At present, \add{investigation of metal-atom movements} in molecular junctions rely mostly on STM or break-junctions in the regime of quantum point contact \cite{tsutsui_atomistic_2009, roslawska_atomic-scale_2021, roslawska_mapping_2022, liu_inelastic_2023}. We demonstrate that a bulk nanoparticle-electrode system can capture 
\add{fluctuations of atom-molecular contacts} in both the conductance and light emission and extend the optoelectronic investigation of quantum-point contacts in STM/break-junctions to a completely different regime of conductance, junction geometry, and operation conditions. 

\section*{Results and discussion}

\begin{figure}[H]
		\centering
		\includegraphics[width=\textwidth]{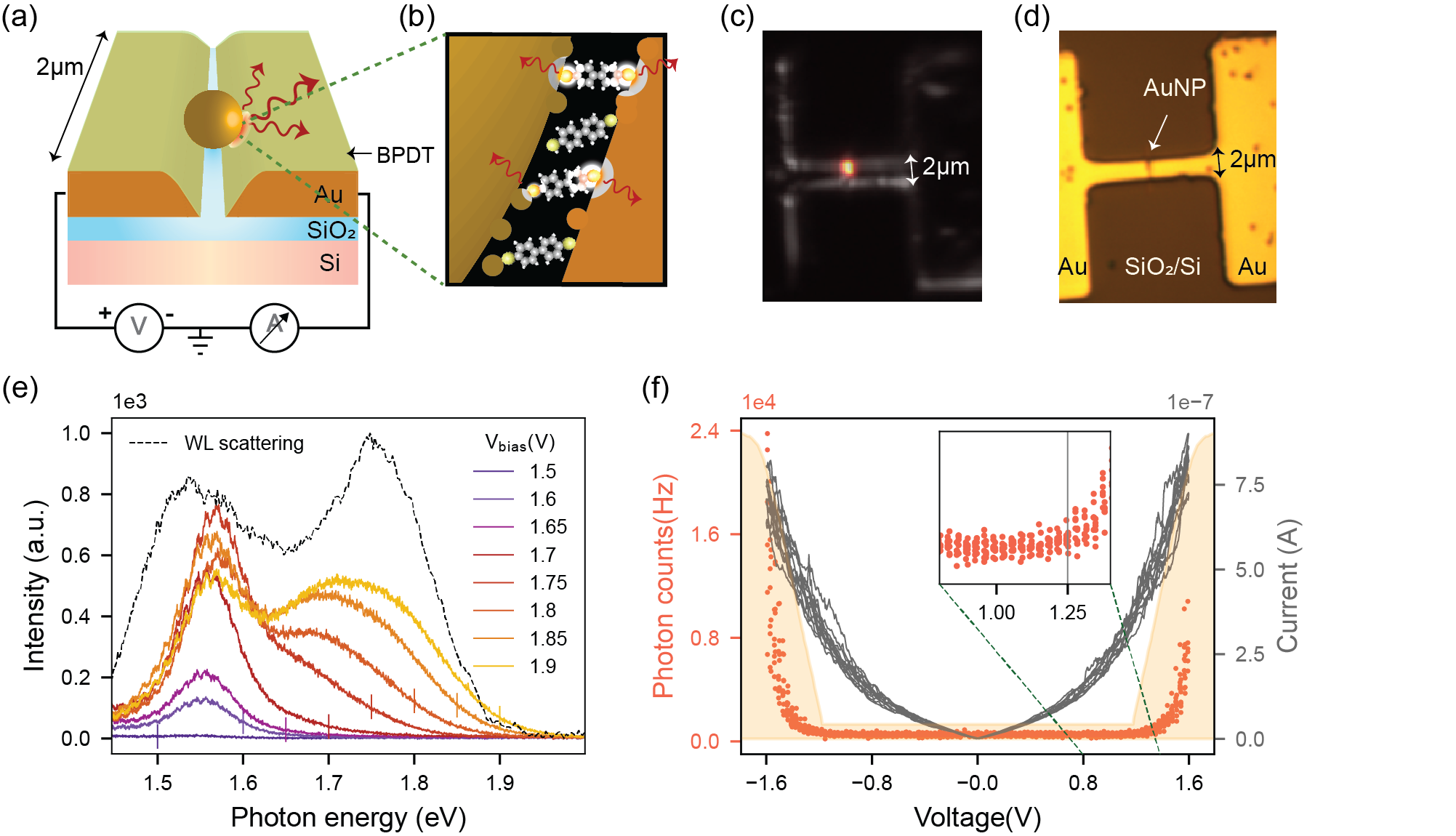}
		\caption{(a) Schematic illustration of the PMJ. (b) Illustration of atomic fluctuations and light emission at the metal-molecule-metal contacts. (c) Grayscale dark field image overlaid with the colored light emission image under 1.6~V bias for a typical PMJ. (d) Corresponding bright field image. (e) Light emission spectrum for various DC voltages with 20~s exposure time. The vertical lines mark the onset of overbias emission. Dotted lines represent the dark-field scattering spectrum from a white light source. (f) Current vs. voltage characteristics of the junction (gray curve) and simultaneously collected photon counts (red curve) as a function of applied voltage. The shaded background represents the quantum efficiency of the detector with voltage values translated to photon energy in eV. Inset: zoomed-in view with the vertical line showing the onset of photon detection.}
  \label{fig:scheme}
\end{figure}

The plasmonic molecular junction (PMJ) is formed by gold nanoparticles (from a citrate-stabilized colloidal suspension) bridging two gold electrodes that are previously functionalized with a SAM of biphenyl-4,4'-dithiol (BPDT) molecules (Fig.~\ref{fig:scheme}a). One or few gold nanoparticles are trapped in the gap and establish electrical contact (see Supporting Information Sec.~\ref{secSI:Experimental methods} for sample fabrication and experimental methods). Each nanoparticle linker forms in fact two junctions in series, but one of them typically dominates the series resistance and experiences most of the voltage drop, as will be discussed below. \add{The device conductance $G$ after fabrication falls in one of the following ranges: (1) Short-circuited contact where $G\sim 10^2\cdot G_0$ where  $G_0 = 2e^2/h$ is the quantum of conductance; this occurs in particular in junctions fused by ESD. (2) Open-circuited contact where $G<10^{-6}\cdot G_0$; this occurs due to several reasons: (i) there is no nanoparticle bridging the gap; (ii) the size of the nanoparticle is smaller than the size of the gap; (iii) the nanoparticle does not establish successful contact with the electrodes on both sides. (3) Molecular contact where $10^{-5} \cdot G_0<G<10^{-1}\cdot G_0$. Sometimes an initially open-circuited junction can be brought into molecular contact by applying several volts.} We also note that junctions featuring the largest conductances ($G>2\times10^{-3\cdot} G_0$) cannot be studied for light emission, as under voltages above 1.2~V, the corresponding current exceeds the damage threshold of the device, which is typically a few $\mu$A and above which irreversible changes occur (see Supporting Information Sec.~\ref{secSI:Light from IET}).  

The nanogaps formed by the SAM between the metallic parts have a width of around 1~nm that depends on the length of the molecules and their orientation with respect to the gold surfaces \cite{ahmed_structural_2021}. They support localized plasmonic resonances that offer extreme light confinement and good radiative efficiency \cite{baumberg_extreme_2019} and allow to efficiently read out the plasmonically enhanced optical signals.  
When a voltage bias is applied across this device, the PMJ emits light originating from the optical-frequency shot noise of electrons inelastically tunneling across the junction through the potential barrier \add{as discussed below} \cite{lambe_light_1976, hone_theory_1978, schneider_optical_2010, kern_electrically_2015, fevrier_tunneling_2018}. An image of light emission from one PMJ overlaid with a dark-field scattering image is shown in Fig.~\ref{fig:scheme}c, with the corresponding bright-field image shown in Fig.~\ref{fig:scheme}d. The emission couples to the plasmonic modes of the cavities formed by the nanogaps between the nanoparticles and electrodes and can be observed in the far field.

The light emission spectrum of BPDT junctions for different d.c. bias voltages and an exposure time of 20~s is shown in Fig.~\ref{fig:scheme}e. \add{Note that there is a cut-off observed in the light emission spectrum at an energy close to the value of $e$V (V is the applied voltage), as indicated by the vertical line in Fig.~\ref{fig:scheme}e. This is also asserted in Fig.~\ref{fig:scheme}f, where the photon count rate increases nonlinearly beyond a threshold of about 1.25~V that matches the spectral response of the silicon detector having a cut-on around this photon energy.} While we expected the electrode-molecule-nanoparticle geometry to form two identical molecular junctions in series, the fact that the energy cut-off matches well with the applied voltage suggests that most of the voltage drops across one of the two junctions (consistent with previous observations of similar junctions \cite{kern_electrically_2015}) and the light emission happens through a single-electron inelastic tunneling process. 

Some overbiased light emission is also observed in our PMJs, with photon energies higher than the applied voltage (times the electron charge). This was previously demonstrated in STM-junctions\cite{schull_electron-plasmon_2009, xu_overbias_2014, kaasbjerg_theory_2015,peters_quantum_2017, fung_too_2020}, electromigrated junctions \cite{downes_measurement_2002, cui_electrically_2020, zhu_tuning_2022}, mechanical break junctions\cite{malinowski_infrared_2016} and more recently memristive junctions\cite{hamdad_overbias_2022}. This emission was either attributed to hot carriers or coherent multi-electron scattering processes. In our experiment, the excess photon energy is in agreement with a moderate increase of the electron temperature in the junction \cite{buret_spontaneous_2015} (see Sec.~\ref{secSI:Thermal emission} in Supporting Information).

The two peaks in the spectra are attributed to plasmonic modes of the junction, also seen in the dark-field scattering spectrum (dotted lines) collected using a tungsten lamp. \add{The nanoparticle-SAM-electrode design is adapted from a previous work involving a nanoparticle-in-groove dual resonant cavity\cite{chen_continuous-wave_2021} that exhibits several resonances at visible and NIR wavelengths.}
\add{There is no evidence of hot electron emission in our system (see Sec.~\ref{secSI:Thermal emission}) and the emission is found to be shaped by the plasmonic response as discussed in Sec.~\ref{SI_sec:pol_response} of Supporting Information}. \add{We refer to Ref.~\cite{chen_continuous-wave_2021} for simulation of the visible plasmonic modes of a similar structure.} The plasmonic modes are progressively populated upon increasing the voltage, as observed with the onset of emission from a plasmonic mode at higher energy for voltages above 1.7~V. 

\begin{figure}[H]
	\centering
	\includegraphics[width=\textwidth]{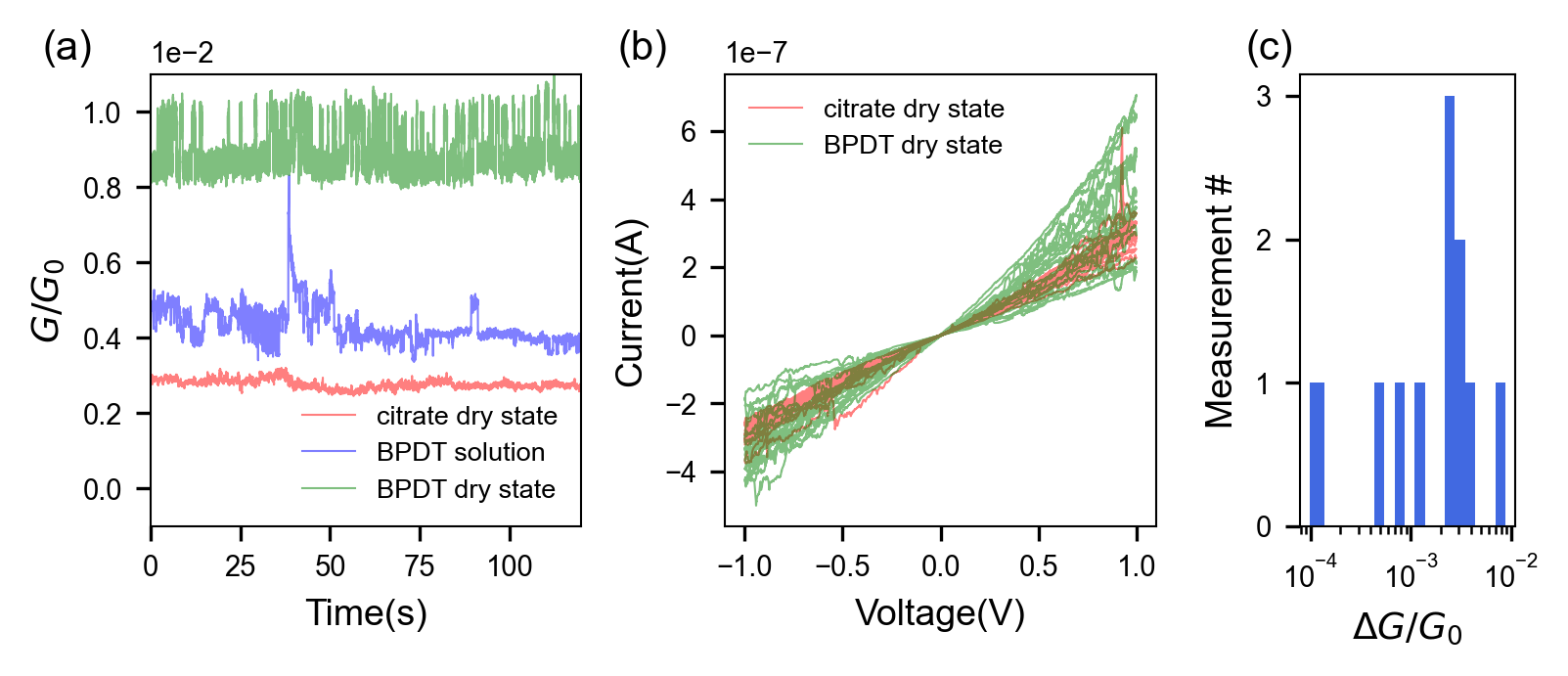}
        \caption{(a) Conductance of a particular PMJ with citrate (red) spacer in a dry state, during deposition of BPDT molecules (blue) and after deposition of the BPDT molecules (green). (b) Current-voltage characteristics of the PMJ before and after deposition of BPDT molecules. \add{ (c) Magnitude of conductance changes evaluated from several measurements where clear intermittent blinking was observed. The data is collected from 6 distinct devices(see Fig.~\ref{figSI:intermittent_cond} in Supporting Information).} }
	\label{fig:BPDT citrate PMJ}
\end{figure}

\begin{figure}[H]
		\centering
		\includegraphics[width=\textwidth]{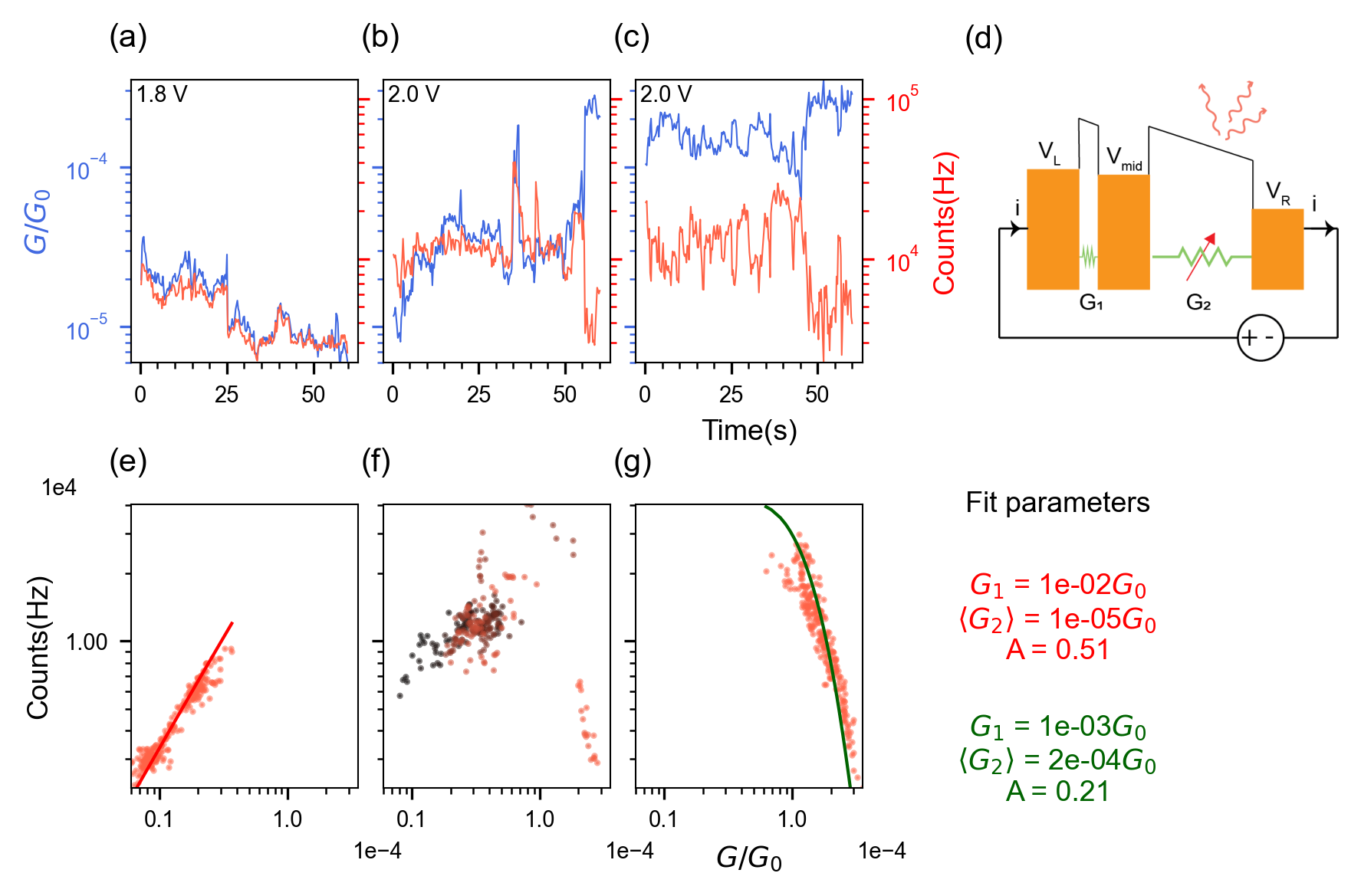}
		\caption{(a-c) Conductance (blue lines) and photon counts simultaneously measured with SPCM (red lines) for a particular PMJ. Both data are summed into 500~ms time bins to improve the signal-to-noise ratio. (d) Schematic representation of the device with fluctuating conductance model. (e-g) Corresponding correlation plots of the data from (a-c) display the switching between positive and negative correlations (see transition within (f) - the data points are color-coded from black to red, representing the progression of time). The solid lines in (e) and (g) are from the model discussed in the text, with the corresponding parameters indicated on the right panel.}
  \label{fig:spcm-timeseries}
\end{figure}

\begin{figure}[H]
		\centering
		\includegraphics[width=1\textwidth]{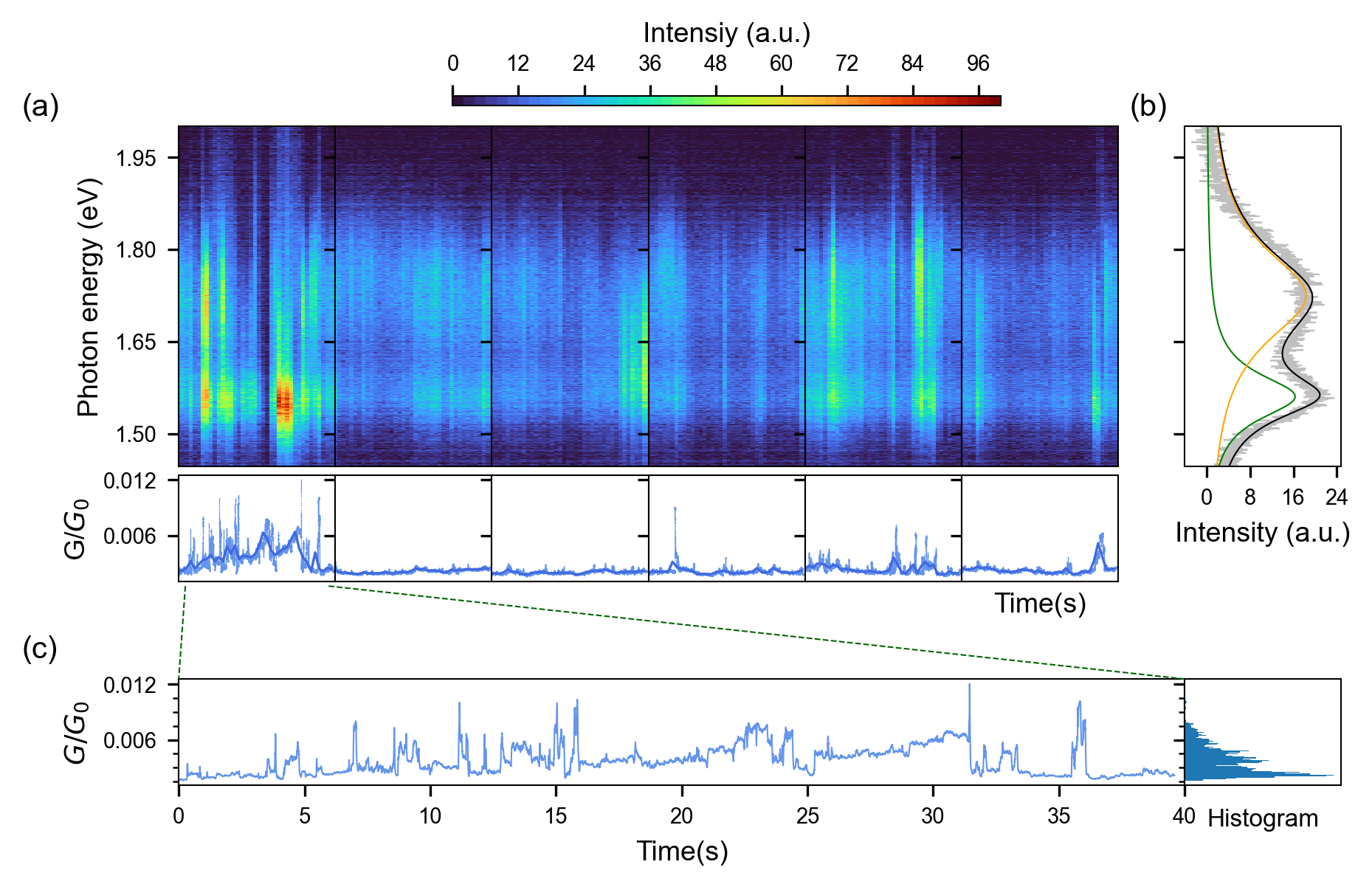}
		\caption{(a) Atomic fluctuations captured in the light emission spectra (top panel) and the electrical conductance $G/G_0$ (bottom panel) collected simultaneously from the same PMJ. A constant DC bias voltage of 1.9~V was applied, spectra were collected with 1~s exposure time and the conductance data were measured simultaneously at a 2~kHz sampling rate. (b) The grey curve represents the average of all the spectra in (a), and the black solid line illustrates the fit obtained from two Lorentzian peaks (orange and green lines). (c) A subset of electrical transport data from (a) showing discrete jumps in the conductance. The histogram of the conductance data is shown in the right panel, suggesting a quasi-continuum of conductance states.}
  \label{fig:wandering}
\end{figure}

Despite using SAMs as molecular-ensemble contacts, we typically find conductance values that are rather in the range of few-molecule junctions \cite{mishchenko_influence_2010, kos_quantum_2021}. We, therefore, expect that the resulting light emission from IET originates from one or few spatially localized emitters (i.e., point sources) whose localization is varying over time across the electrode gap, as pictured in Fig.~\ref{fig:scheme}b. We now study the direct optical and electrical signatures associated with the dynamic localization of these point-like emitters as discussed in (Figs.~\ref{fig:BPDT citrate PMJ}-\ref{fig:wandering}).

The current-voltage characteristics (grey curve in Fig.~\ref{fig:scheme}f) show the typical nonlinear electrical response of a molecular junction\cite{gehring_single-molecule_2019}. Given the nanometer size of the molecular spacer, we propose that the main transport mechanism is tunneling. The thiol anchor group of the BPDT molecules forms strong covalent bonds with the gold atoms and the current flow should be mainly mediated by the HOMO levels \cite{pauly_density-functional_2008}. However, since we perform the experiment at room temperature the resonant molecular orbitals of BPDT cannot be identified from the current-voltage characteristics. Also, note that the HOMO-LUMO gap of BPDT molecules is high (3.85~eV\cite{pauly_density-functional_2008}) and hence de-excitation through the molecular orbitals is not involved in the light emission process for the voltages considered in this study. There could be additional tunneling contribution from the citrate molecules present in the gap as BPDT may not replace all the citrate molecules around the nanoparticles. However, the citrate molecules are weakly bonded to the gold atoms and form an even smaller nanogap between the electrodes\cite{govor_conductance_2010}. Pure citrate junctions are characterized in Supporting Information (Sec.~\ref{secSI:Yield of the devices}) which show a much lower yield of conductive junctions compared to the BPDT junctions. The presence of BPDT molecules on the surface of gold readily attracts nanoparticles in the nanogap to bridge the electrodes and establish electrical contact. 

In order to further elucidate the nature of the spacers in the PMJ, a progressive addition of BPDT to a native PMJ with a pure citrate spacer is carried out. A PMJ is first formed by depositing citrate-capped nanoparticles in the electrode gap. A solution of BPDT molecules is dropped on the PMJ and the conductance is monitored simultaneously. We observe that conductance fluctuations become more pronounced when the solution of BPDT is present. After the droplet evaporates, some BPDT molecules are expected to replace the citrate spacer in the nanogap. We find that the D.C. conductance (Fig.~\ref{fig:BPDT citrate PMJ}a) and the current-voltage characteristics (Fig.~\ref{fig:BPDT citrate PMJ}b) of the PMJ with BPDT fluctuates more than the initial PMJ with citrate spacer (see also \add{Sec.~\ref{secSI:devices-IV}} of Supporting Information). The increased fluctuation of current density in BPDT junction is attributed to the formation of gold-thiol `staples'. The gold-thiol bond is capable of lifting adatoms on metal surfaces \cite{burgi_properties_2015, benz_single-molecule_2016, carnegie_room-temperature_2018} \add{and were found to increase the frequency of these events.} 

Apart from an increase in the average conductance value post-functionalization (that may further evolve over long measurement times, see Supporting Information Sec.~\ref{secSI:Stability of the devices}), we observe clear telegraph noise that is consistent with a randomly switching contact at the single-molecule level (Fig.~\ref{fig:BPDT citrate PMJ}a). 
The biphenyl molecule's conductance depends on its conformation \cite{mishchenko_influence_2010} and coordination geometry of the terminal thiols to the gold atoms \cite{burkle_conduction_2012}. However, the changes in conductance due to these intramolecular phenomena are averaged out at room temperature where the aromatic systems are oscillating around an average position. The large discrete conductance jumps, which can reach an order of magnitude in some cases, are mainly attributed to the binding and unbinding of metal-molecule contacts \cite{brunner_random_2014}, the time-scale of which has been found to be on the order of a second at room temperature \cite{carnegie_room-temperature_2018}. \add{Post-selecting conductance measurements from 6 distinct BPDT PMJs at a bias voltage of 5~mV that show clear intermittent blinking, the magnitude of the conductance jump is estimated to be between $1\times10^{-4}\;G_0$ and $8.7\times10^{-3}\;G_0$ (Fig.~\ref{fig:BPDT citrate PMJ}c), which is in agreement with the conductance of a single BPDT molecule found in literature \cite{guo_measurement_2011,jeong_-chip_2020,ramachandran_electromechanical_2018,domulevicz_multidimensional_2021}. The conductance variation due to conformational change is much lower in magnitude \cite{vonlanthen_chemically_2009, mishchenko_influence_2010}.} Breaking of a single molecular contact mostly occurs at the Au-Au bond which has a bond strength of 1~eV as compared to the Au-S bond strength of 1.6~eV \cite{tsutsui_quantitative_2009}. Thus, measuring the conductance of the PMJ provides information about atomic fluctuations at the metal-molecule interface.
The point-like IET emitters resemble randomly occurring adatom protrusions at metallic nanogaps that can locally affect the plasmonic response.  \add{These atomic-scale events} are usually probed by laser spectroscopy, mainly by SERS\cite{benz_single-molecule_2016} and TERS\cite{liu_dramatic_2020} and more recently manipulated by STM in the regime of quantum point contact \cite{roslawska_atomic-scale_2021, roslawska_mapping_2022, liu_inelastic_2023}. Here, we show a novel probe of \add{atomic-scale fluctuations} driven by electrical current in a robust plasmonic molecular junction: our data are consistent with the picture of randomly occurring adatoms that create preferential sites for conductance and light emission. 

Our PMJs remain functional over weeks and months (see Sec.~\ref{secSI:Stability of the devices} in Supporting Information), allowing us to perform in-depth characterizations of their dynamic behavior. 
In particular, we can correlate the electrical properties with optical signals associated with the fluctuating nature of metal-molecule contacts, as illustrated in Fig.~\ref{fig:spcm-timeseries}.
We investigate these fluctuations under a constant bias using a single photon counting module (SPCM) for faster acquisition synchronized with the conductance measurement. \add{
Our general finding is that light emission intensity and conductance value are strongly correlated despite the relatively large area of the junction and the presence of hundreds of molecules. It is compatible with the existence of a few molecular sites that dominate both transport and light emission, and whose number fluctuates over time.  
For the same junction, the conductance-emission correlation can switch between positive and negative trends over extended measurement times, as illustrated in Fig.~\ref{fig:spcm-timeseries}e-g (additional data sets are provided in Supporting Information Sec.~\ref{SI_Sec:Additional examples}). The light emission from the IET process is, however, expected to vary linearly with current \cite{sivel_interpretation_1995} and to display abrupt changes only when switching from tunneling to quantum point contact \cite{schull_electron-plasmon_2009, schneider_optical_2010}. The deviation from linearity and the periods of negative correlation observed in our PMJs will now be explained by the interplay between two nanojunctions in series.} 

As mentioned earlier most of the voltage drop happens across one of the two junctions in series and the light emission is expected to originate from this dominating one as illustrated in Fig.~\ref{fig:spcm-timeseries}d. Hence, the conductance $G_1$ of the non-emitting junction is assumed to be much greater than $G_2$, that of the light emitting junction. Consequently, small fluctuations in \add{$G_1$} have insignificant impacts on overall conductance and emission intensity.
The instantaneous light emission intensity depends on three major components: (i) the current through the emitting junction \add{$i(t)$} \cite{schull_electron-plasmon_2009}; (ii) the voltage across the emitting junction $V_{mid} - V_R$ (through the energy-dependent responsivity of the detector); and (iii) the efficiency for converting electrical energy into photons, which varies with the emitter location \cite{chen_intrinsic_2021} (see Fig.~\ref{fig:wandering}a,b). We will show that contributions (i) and (ii) can together explain both negative and positive correlations of photon counts vs. conductance observed in the same junction at different times, while (iii) results in a variation of the overall efficiency of the same PMJ accounted for by the prefactor $A$ below. 
The intensity of light emission is therefore modeled by the equation
\begin{equation}
    I_{PMJ}(t) = A\:.i(t)\:.f(V(t))
    \label{eq:photon-fit}
\end{equation}
where $f(V(t))$ is an experimentally determined function that accounts for the time-averaged voltage-dependent emission and detection efficiency. The linear dependence on $i(t)$ is to be interpreted as a first-order Taylor expansion valid for a limited range of conductance fluctuations. 
From the conductance data collected at constant bias, we fit the data from Fig.~\ref{fig:spcm-timeseries} to obtain the values of the parameters $G_1$ and $G_2(t)$. These parameters are used to derive the values of $V_{mid} - V_R$ for each value of the conductance. 

The empirical relation from equation~\ref{eq:photon-fit} is then used to obtain the fit shown as solid lines in Fig.~\ref{fig:spcm-timeseries}e,g. When the change in $V_{mid} - V_R$ is negligible, the monotonous dependence of photoemission on the current dominates and leads to positive correlations between photon count rate and conductance. This can happen when $G_1>>G_2$ such that $V_{L} - V_{mid}$ approaches 0. This is the case for the fit parameters $G_1 = 10^{-2}\;G_0$ and $\langle G_2 \rangle=10^{-5}\;G_0$ in Fig.~\ref{fig:spcm-timeseries}e (the notation $\langle G_i \rangle$ is used to represent the mean value of the conductance of junction $i$=1,2). Conversely, negative correlations are observed when $V_{mid} - V_R$ changes significantly so that the voltage dependence of emission embodied in $f(V(t))$ dominates over the dependence on current. 
This is the case for fit parameters $G_1 = 10^{-3}\;G_0$ and $\langle G_2 \rangle=2\times10^{-4}\;G_0$ in Fig.~\ref{fig:spcm-timeseries}g. Depending on the values of $G_1$ and $G_2(t)$ the two opposite regimes can be realized. In all fits, $G_1$ > $G_2$ as predicted, consistent with our initial assumption that one of the two junctions has a higher conductance compared to the other one. More details on the data fitting are provided in Supporting Information Sec.~\ref{secSI:Details of the fit} and more examples of measured and fitted correlations are presented in Sec.~\ref{SI_Sec:Additional examples}.

In Fig.~\ref{fig:wandering} we report the fluctuations in light emission spectrum occuring together with conductance fluctuations in the same PMJ. The light emission spectrum is dominated by two peaks at photon energies of 1.56~eV and 1.72~eV (as estimated by the Lorentzian fits in Fig.~\ref{fig:wandering}b), but their respective intensities can randomly change over time. 
\add{Intermittent blinking in single-molecule junctions can happen from conformational or structural changes in the molecular backbone. But such effects take place in timescales of $10^{-11}$ to $10^{-9}s$ and are averaged out in large area molecular junction \cite{jones_single_2007, paulsson_conductance_2009, brunner_random_2014}. A gold-SAM-EGaIn junction with a small contact area was able to capture such intermittent blinking in light emission but found no correlation with current fluctuations \cite{du_-chip_2016, wang_operando_2019}. The BPDT molecules can display conductance switching by a change in the tilt of the molecule in the junction, its binding conformation with the gold atoms, and the rotation between the two benzene molecules \cite{mishchenko_influence_2010, burkle_conduction_2012}. These effects could explain blinking in the intensity of the light emission due to the change in conductance \cite{kos_optical_2020} but do not explain the reshaping of the plasmonic modes.}

Instead, we argue that \add{dynamically occuring} localized  \add{atom-molecular contacts} naturally explain the observed fluctuations in the light emission spectrum. 
\add{We hypothesize that the changing ratio between emission intensities at the different plasmonic peaks is caused by the random appearance of point-like emitters at different positions inside the host nanocavity. Depending on its localization, a point-like emitter couples more efficiently to a particular mode of the plasmonic response.  The far-field spectra of these point-like emitters depend on the overlap between the emitter position and the near-field distribution of the different gap modes. Hence, the \textit{spatial} wandering of point-like emitters in the near-field causes an apparent \textit{spectral} wandering in the far-field. This has been demonstrated in photo-excited luminescence of gold clusters in a plasmonic cavity by Chen \textit{ et al.} in Ref.~\cite{chen_intrinsic_2021}, where simulation of this effect in a gap nanocavity similar to the ones studied here was performed.}

The relative conductance fluctuations in our PMJs are found to be significantly enhanced under increasing electric bias (see Supporting Information Sec.~\ref{secSI:voltage induced fluatuations}) while there is no clear temperature rise seen in the overbias emission, within the fitting uncertainty (Fig.~\ref{figSI:overbias}). These observations suggest that a non-thermal mechanism may contribute to the creation of new atomic protrusions and could be connected to recent findings on optically-induced picocavities \cite{lin_optical_2022}, where external electric fields (at the optical frequency) were argued to lower the energy barrier for the creation and relaxation of a gold adatom. However, future dedicated experiments are needed to confirm the non-thermal mechanism that may be at play in the junction.

\section*{Conclusion}
In summary, using a simple and \add{scalable} self-assembled geometry, we demonstrated how the structural rearrangements of atom-molecule contacts in a plasmonic molecular nanojunction are imprinted on the electrical transport as well as the tunneling-induced light emission fluctuations. 
The non-monotonous correlations between electrical conductance and IET intensity are explained by a simple phenomenological model taking into account that our devices consist of a double junction in series. \add{The large intermittent fluctuations in conductance and IET emission are proposed to arise from the binding and unbinding of molecular contacts due to the movement of gold adatoms; the exact location of these dominant conduction channels governs the relative coupling of IET to distinct plasmonic modes.} While the \add{switching of atom-molecule contacts} is not a deterministic process, we could repeatedly drive our system in this regime by applying a sufficient voltage across the junction.

With controlled capillary assembly \cite{flauraud_nanoscale_2017}, atomic force microscope \cite{kern_electrically_2015}, transfer printing \cite{redolat_accurate_2023} or dielectrophoresis \cite{yoon_dielectrophoretic_2008}, it is possible to make on-demand single nanoparticle junctions (some preliminary results are shown in Sec.~\ref{secSI: DEP} of Supporting Information). This would enable simultaneous SERS or luminescence and electrical measurements on the PMJ, which are currently hindered by the presence of multiple nanoparticles nearby. The resulting single-nanoparticle PMJ offers a unique opportunity to connect the microscopic origins of various phenomena such as picocavity in SERS \cite{benz_single-molecule_2016}, flares in electronic Raman scattering \cite{carnegie_flickering_2020}, blinking of gold photoluminescence \cite{chen_intrinsic_2021}, and flucutations in IET and conductance studied here.
\add{The PMJs should be particularly useful in understanding their formation mechanisms, including their dependence on the electric field \cite{lin_optical_2022} and their non-thermal origin. }

\section*{Funding}

This work received funding from the European Union’s Horizon 2020 research and innovation program under Grant Agreement No. 820196 (ERC CoG QTONE).
C.G. acknowledges the support from the Swiss National Science Foundation (project numbers 170684 and 198898).
A.B. received funding from the French Agence Nationale de la Recherche (ANR-20-CE24-0001 DALHAI, ANR-21-ESRE-0040 Smartlight, ISITE-BFC ANR-15-IDEX-0003, ANR-17-EURE-0002 EIPHI Graduate School), from the Région de Bourgogne Franche-Comté, from the European Regional Development Fund (FEDER-FSE Bourgogne Franche-Comté 2021/2027), and the CNRS.

\newpage

\section{Experimental methods}
\label{secSI:Experimental methods}
\subsection{Sample fabrication}

\begin{figure}[H]
		\centering
		\includegraphics[width=0.9\textwidth]{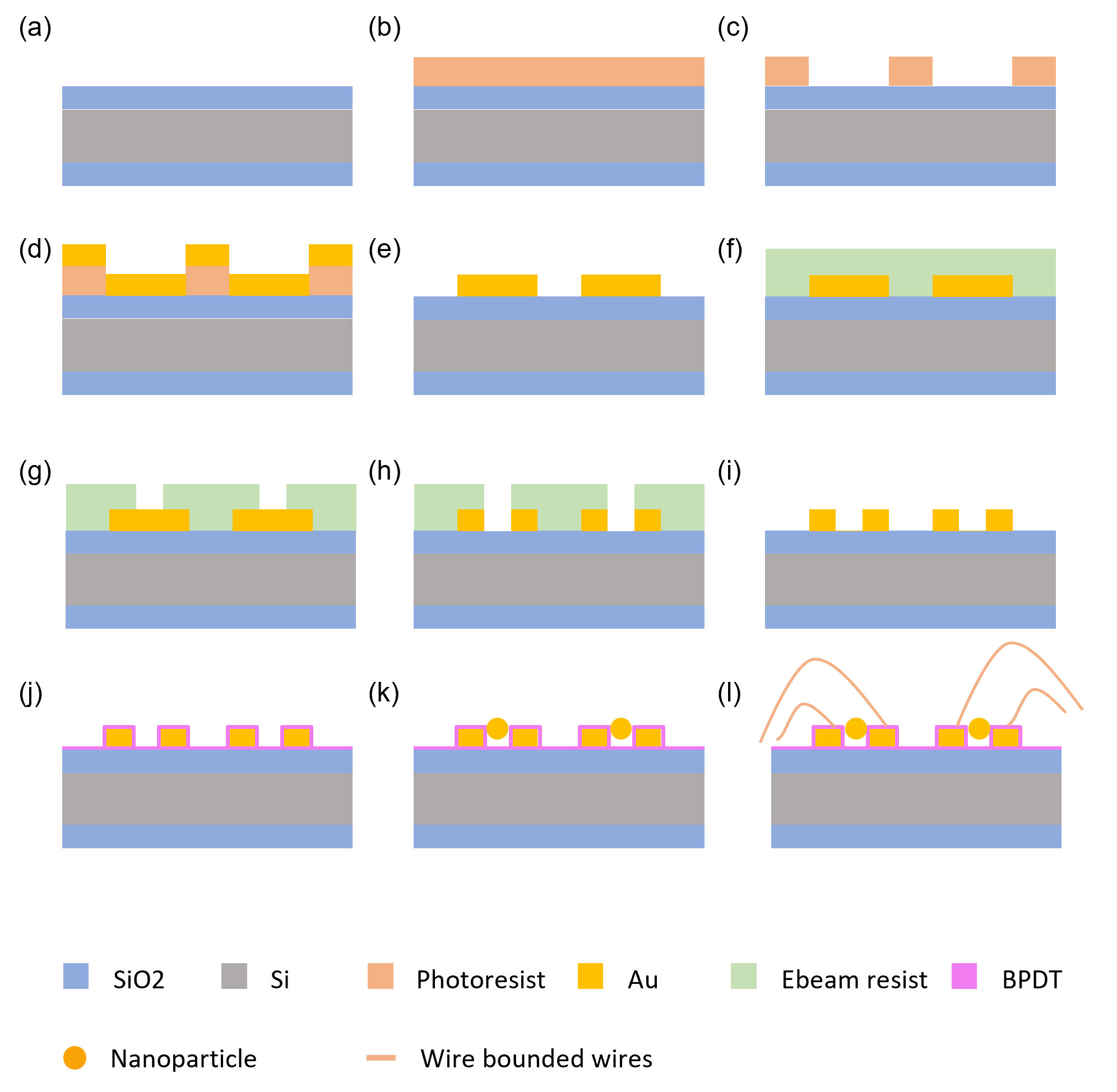}
        \caption{Fabrication steps: (a) Substrate with 380~$\mu$m Si and double-sided 300~nm SiO2 layers. (b) Spin coating with 700~nm LOR~5A and 1600~nm of AZ1512. (c) Deep-UV (DUV) photolithography. (d) Thermal evaporation of a 3~nm Cr adhesion layer followed by 150~nm of Au. (e) Lift-off process. (f) Spin coating with E-beam resist 950K. (g) Electron-beam lithography. (h) Ion-beam etching. (i) Removal of E-beam resist. (j) Formation of molecular spacer. (k) Nanoparticle drop casting. (l) Wire bonding.}
  \label{figSI:fabrication}
\end{figure}

The sample fabrication procedure is illustrated in Fig.~\ref{figSI:fabrication}. A 4-inch diameter silicon wafer with a thickness of 380~$\mu$m and double-sided 300~nm SiO2 layers (NOVA wafers) is used as the substrate for the fabrication process. The substrate underwent a spin coating process, with a 700~nm layer of LOR~5A followed by a 1600~nm layer of AZ1512 photoresist.

A first photo-lithography is employed to define the contacts and gold electrodes by metal deposition and lift-off.
Utilizing the deep-UV photolithography technique, the electrode pattern is formed on the substrate. Following the photoresist development step, a 3-nm-thick Cr adhesion layer and a 150-nm-thick gold layer are thermally evaporated onto the substrate at a rate of 0.5~nm/s. The subsequent metal lift-off process is performed using Remover 1165~-~NMP.

 Subsequently, electron-beam lithography and collimated ion-beam etching are employed to carve the $\sim 150$ nm gaps separating the two electrodes, creating the space for the bridging gold nanoparticles to fit.
We used 100~keV electron-beam lithography and PMMA 950~K as the e-beam resist. Following the development of the resist, the nanogaps are etched with a collimated ion-beam at an angle of -10$^{\circ}$, resulting in a 150~nm wide and 4~$\mu$m long V-shape trenches, oriented perpendicular to each electrode. The actual length of the gold electrode is 2~$\mu$m; the e-beam step creates a longer gap that extends to the substrate to safely mitigate alignment errors.
To remove PMMA, the sample is immersed in warm Remover 1165~-~NMP for a duration of 6 hours.

After the fabrication of electrodes, the wafer is diced into individual chips, with each chip size (7~mm x 10~mm) containing 25 pairs of electrodes. Following the dicing process, the chips underwent the formation of molecular spacer and nanoparticle bridges, as described in section \ref{secSI: Formation of molecular spacer and nanoparticle bridge}. Chips are then forwarded for wire bonding to connect the electrodes to a printed circuit board (PCB). This wire-bonding step enabled the electrical connection between the fabricated structures and the external circuitry, facilitating further characterization with the setup described in section \ref{secSI:Optical setup}.

\subsection{Formation of molecular spacer and nanoparticle bridge}
\label{secSI: Formation of molecular spacer and nanoparticle bridge}
\begin{figure}[H]
		\centering
		\includegraphics[width=0.6\textwidth]{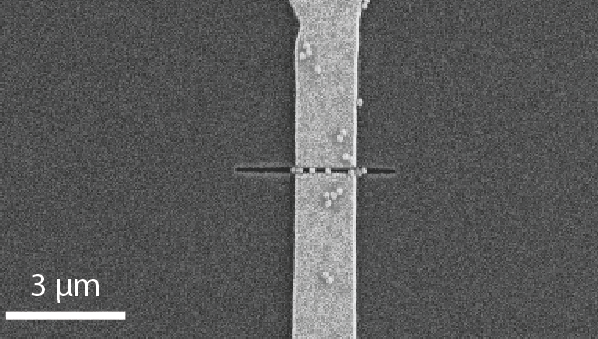}
		\caption{SEM image of a plasmonic molecular junction (PMJ) with multiple nanoparticles in the gap.}
  \label{figSI:SEM}
\end{figure}

Biphenyl-4,4'-dithiol (BPDT) molecules in solid form from Sigma Aldrich are dissolved in ethanol to form a 3~mM solution. Each chip is incubated in 3~ml of this solution for 2 hours \cite{ahmed_structural_2021}. The sample with the self-assembled monolayer (SAM) of BPDT is then cleaned several times with ethanol to remove unbound molecules. To form the nanoparticle bridge, 5~$\mu$l of 1:100 diluted \textit{Nanopartz} 150~nm OD100 gold nanoparticle solution is drop cast on the sample and evaporated until the edge of the droplet passes across the slit. Due to the capillary forces, the edge has a higher concentration of nanoparticles and leaves some nanoparticles in the slit \cite{flauraud_nanoscale_2017}. In the final device, the electrodes with molecular SAM are bridged by a few nanoparticles as shown in the SEM image in Fig.~\ref{figSI:SEM}. 

 We found that instances with a single nanoparticle have a low probability of establishing good contact with the electrodes. Having more nanoparticles increases the probability of making a successful contact. Note that all the discussion in the main text does not rely on having a single particle bridging the electrodes; having more than one simply increases the total number of molecules potentially participating in conductance and light emission by inelastic electron tunneling. Similarly, the different plasmonic modes that enhance and reshape the emission spectrum may equally be associated with distinct nanoparticles, our model remains fully valid. 

\subsection{Experimental setup}
\label{secSI:Optical setup}
\begin{figure}[H]
		\centering
		\includegraphics[width=0.8\textwidth]{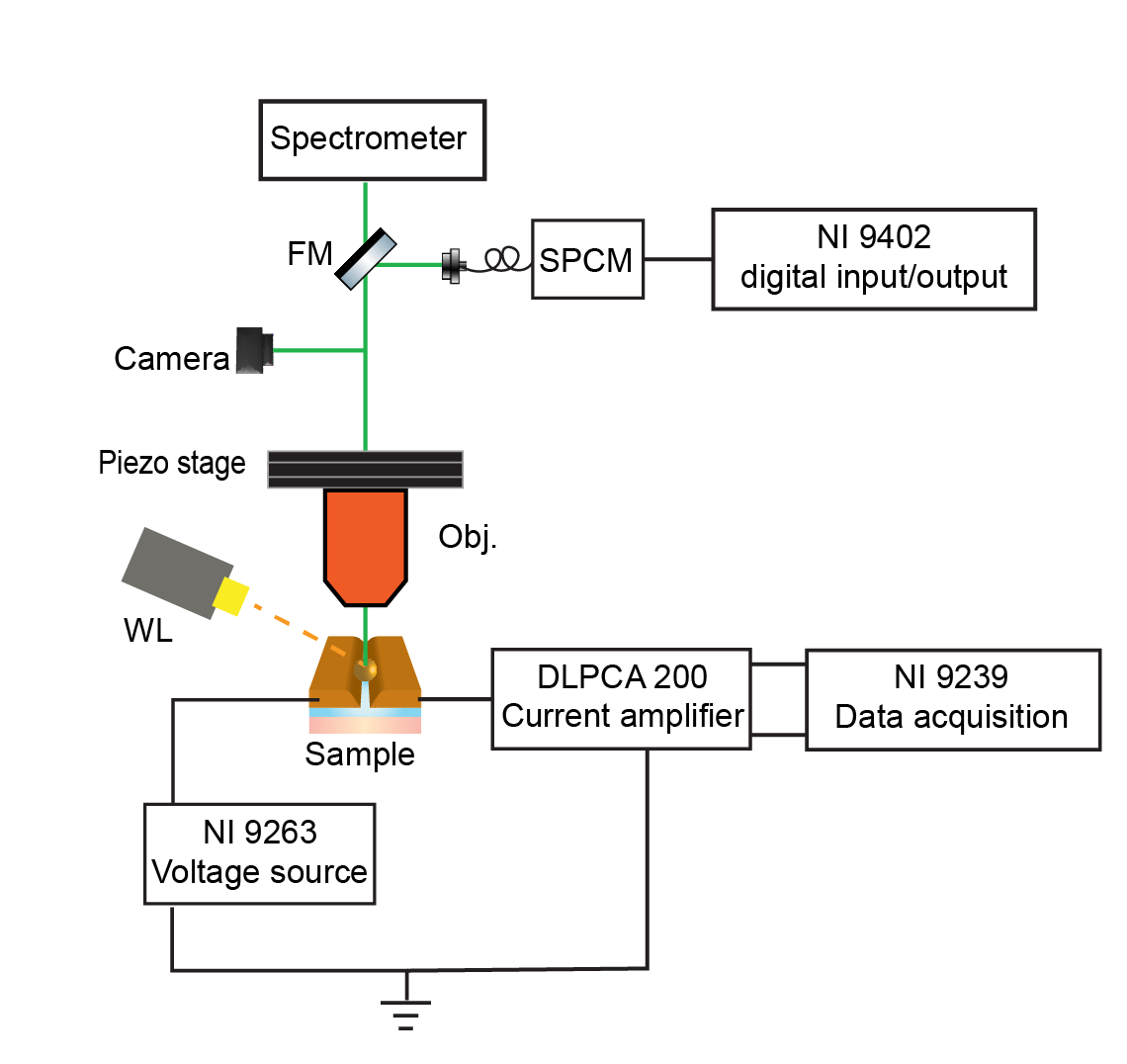}
		\caption{Schematics of the experimental setup: FM- Flip mirror; WL- White light source; Obj.- 0.8NA Objective.}
  \label{figSI:setup}
\end{figure}

The samples are investigated with a home-built optical setup shown in Fig.~\ref{figSI:setup}. The optical signal from the sample is collected using a 0.8~NA objective with 100X magnification and directed to an Andor spectrometer or fiber-coupled to a single photon counting module (SPCM) depending on the measurement. All results discussed in the main text are not normalized to the detection efficiency of the setup. A white light source can illuminate the sample along the laser path or from the side to obtain bright field or dark field imaging, respectively.

\begin{figure}[H]
		\centering
		\includegraphics[height=0.35\textwidth]{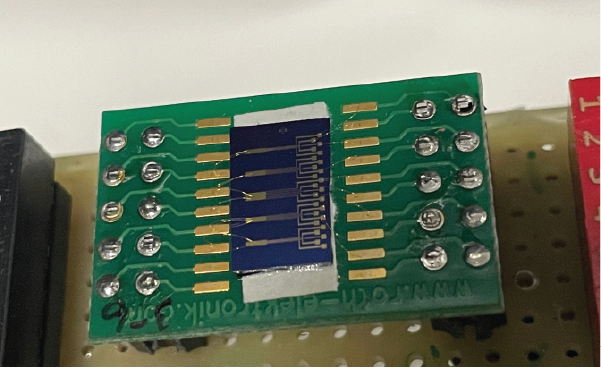}
		\caption{Image of a chip with electrodes wire-bonded to a PCB, which will be further connected to an external electrical circuit.}
  \label{figSI:wirebond}
\end{figure}

To perform electrical measurements, the electrodes are wire-bonded to a PCB (Fig.~\ref{figSI:wirebond}) that is then connected by wires to an external circuit consisting of a voltage source and current measurement unit. The NI~9263 voltage output module can output between -10~V and +10~V. Current-voltage characteristics are measured by applying a triangular voltage sweep. The current is fed to the DLPCA~200 transimpedance amplifier which is automatically gain-switched depending on the magnitude of the measured current by a home-built LABVIEW program. The voltage output of the current amplifier is then read by the NI~9239 analog input device. All the devices are connected to a NI~9040-cRIO controller and are synchronously operated via LABVIEW with an internal clock. Finally, to synchronize the optical measurements with the spectrometer, the NI~9402 pulse generator is used to send a trigger pulse to start the spectral acquisition. The NI~9402 also acts as a digital readout module for photon counting measurements with SPCM.

\section{Intermittent blinking in conductance}
\label{secSI: intermittent blinking condutance}

\begin{figure}[H]
	\centering
	\includegraphics[width=\textwidth]{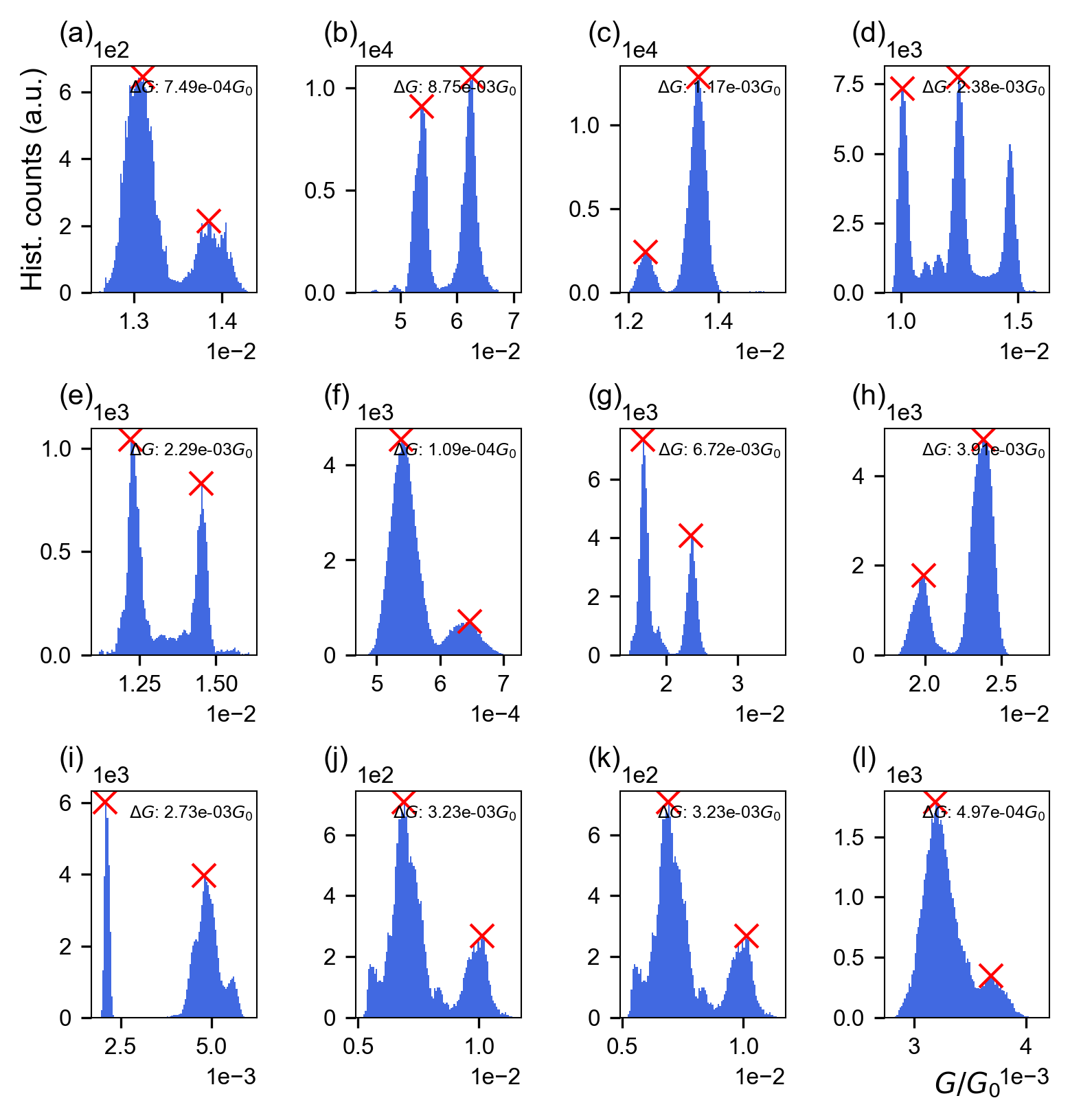}
	\caption{Intermittent blinking in conductance observed in different measurements across different devices at a D.C. voltage of 5~mV}
	\label{figSI:intermittent_cond}
\end{figure}

Intermittent blinking is frequently observed in the conductance measurements of the PMJs. A histogram of conductance values indicates two distinct peaks, as shown in Fig.~\ref{figSI:intermittent_cond} for repeated measurements across 6 different PMJs under a D.C. voltage of 5~mV.
The magnitude of the conductance jump lies between $1\times10^{-4}\;G_0$ and $8.7\times10^{-3}\;G_0$ and is in agreement with the conductance of a single BPDT molecule found in literature \cite{guo_measurement_2011,jeong_-chip_2020,ramachandran_electromechanical_2018,domulevicz_multidimensional_2021}. The conductance change due to conformational change of the molecules is much lower in magnitude \cite{vonlanthen_chemically_2009, mishchenko_influence_2010}. The appearance of three peaks in the histogram (Fig.~\ref{figSI:intermittent_cond}d) that are equally spaced also indicates that it does not originate from conformational changes of the molecules. 

\section{Native ligands vs. BPDT} \label{secSI:citrates_vs_BPDT}

The gold nanoparticles are non-covalently capped with citrate molecules to avoid aggregation of the suspension. When these nanoparticles are deposited in the gap between the electrodes, the BPDT molecules on the surface of the electrodes are expected to replace the citrate layer to form the molecular junction, but some ligand molecules could still be present in the nanogap. To better understand how citrates impact the behavior of the junction, we compare here BPDT junctions with pure citrate-spaced junctions (no incubation in BPDT).

\subsection{Yield of the devices}
\label{secSI:Yield of the devices}
\begin{figure}[H]
	\centering
	\includegraphics[height=0.4\textwidth]{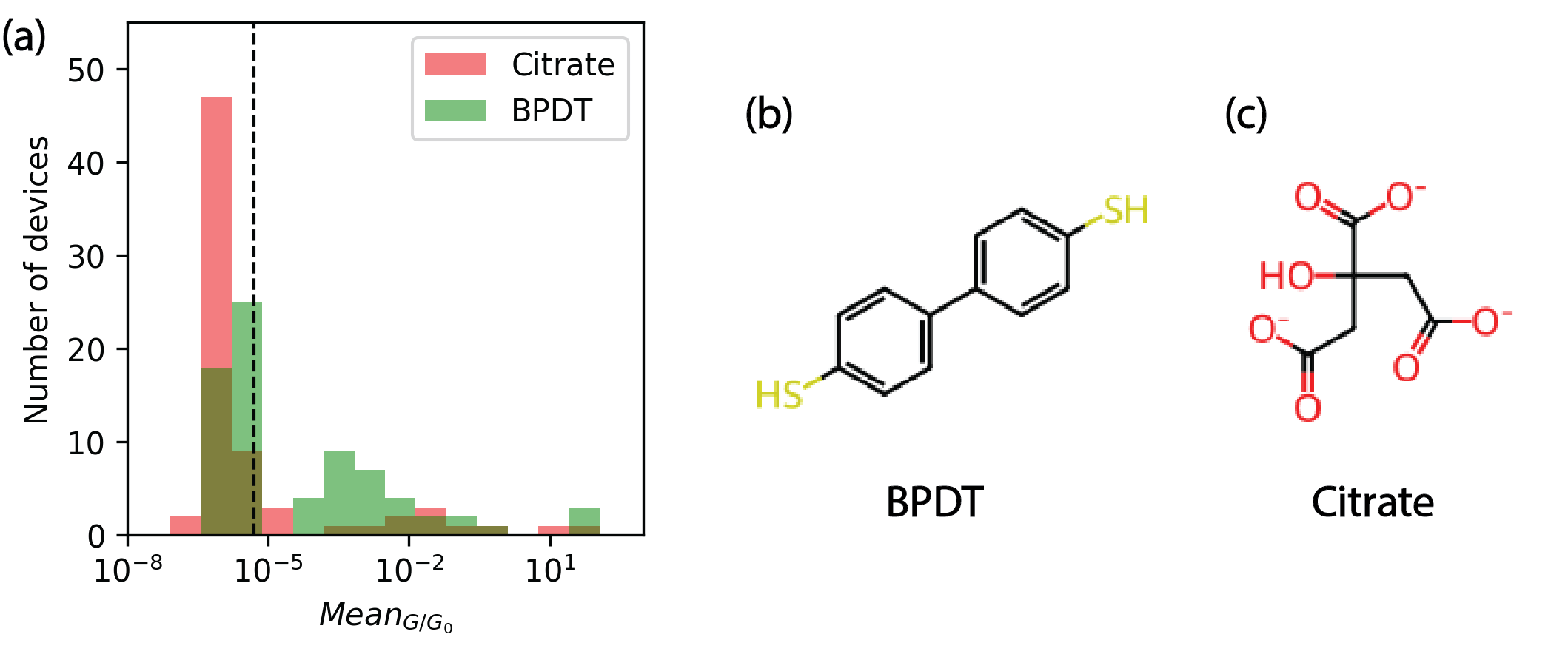}
	\caption{(a) Conductance measured at DC bias of 5~mV for several devices with BPDT (green) and citrate (red) spacer. The dashed line marks the measurement limit of the probe station. BPDT PMJs have a yield of 43\% and the conductance histogram has a peak around $3\times10^{-4} G_0$. Citrate PMJs have a yield of 23\% and the conductance histogram has a peak around $3\times10^{-2} G_0$. Molecular structure of (b) BPDT and (c) citrate molecule}
	\label{figSI:yield}
\end{figure}

Comparing BPDT and pure citrate spacers (2 $\times$ 3 chips, i.e. 75 devices each), the yield of BPDT in establishing measurable electrical contact is 43\% while that of citrate is 23\% (Fig.~\ref{figSI:yield}). The dashed line marks the measurement limit of our probe station, and the devices with conductance below this range are considered open circuits. Considering only the PMJs that we identify as molecular contacts ($10^{-5} G_0<G<10^{-1} G_0$), the histogram peaks at around $3\times10^{-4} G_0$ for the BPDT junction and around $3\times10^{-2} G_0$ for the citrate junctions, which could be related to the smaller size and possibly flatter orientation of the citrate molecules. Here, $G_0 = 2e^2/h$ refers to the quantum of conductance.

\subsection{Current-Voltage characteristics}
\label{secSI:devices-IV}
\begin{figure}[H]
	\centering
	\includegraphics[height=0.4\textwidth]{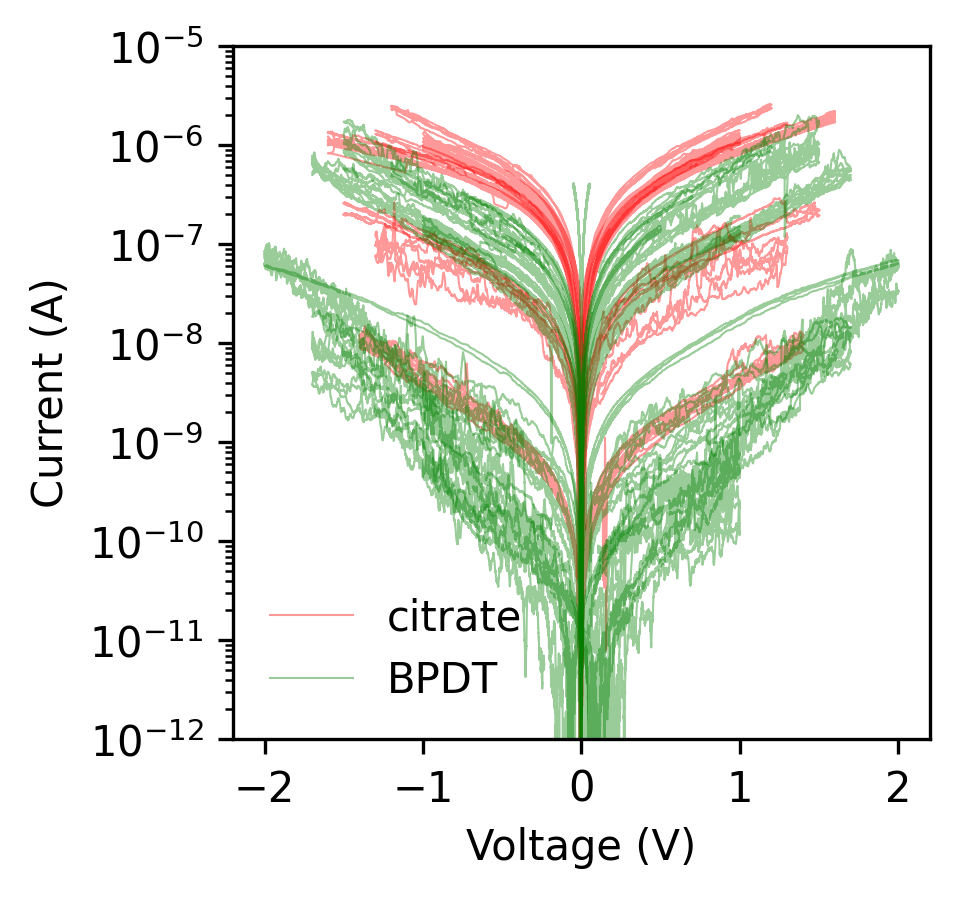}
	\caption{Current-voltage characteristics of several devices with BPDT (green) and citrate (red) spacer.}
	\label{figSI:devices-IV}
\end{figure}

The current-voltage characteristics of several devices with BPDT and citrate spacer are shown in Fig.~\ref{figSI:devices-IV}

A qualitative observation is that PMJs with BPDT spacers are found to feature more pronounced fluctuations in conductance than citrate spacers. However, to clearly distinguish the two kinds of PMJs through their I-V curves, we believe that cryogenic, molecular-specific characterization like inelastic tunneling spectroscopy \cite{jaklevic_molecular_1966} is required.

\subsection{Inelastic electron tunneling (IET) light emission}
\label{secSI:Light from IET}
We observe light emission by IET from both kinds of PMJ (with citrate and BPDT spacers). Our PMJs get damaged when the current through them exceeds a few $\mu$A and hence a current compliance of 1~$\mu$A is maintained. As a consequence, PMJs with higher conductance (above $2\times10^{-3} G_0$ at 5~mV) cannot be used for IET experiments as their currents under the voltages where IET becomes measurable are higher than the current compliance.

\begin{figure}[H]
		\centering
		\includegraphics[width=\textwidth]{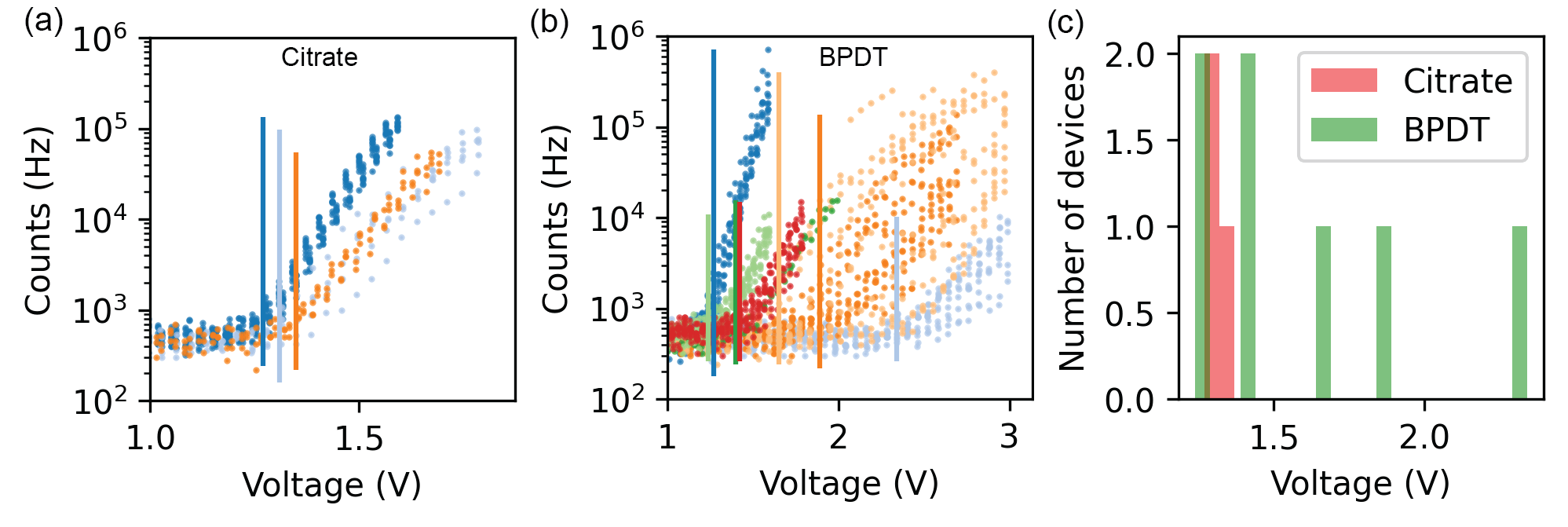}
		\caption{Photon counts versus voltage for several PMJs for (a) citrate spacer and (b) BPDT spacer. Vertical lines mark the voltage threshold for onset of observable IET. (c) Voltage threshold plotted in the form of histogram.}
  \label{figSI:photon-threshold}
\end{figure}

The voltage threshold at which IET is detected is shown for different PMJs with citrate and BPDT spacer (Fig.~\ref{figSI:photon-threshold}). For most devices, the voltage threshold is close to the cut-on energy of the detector efficiency. For other devices with lower conductance, the photon yield is low and they require much higher voltages for IET to be detectable.

\section{Plasmonically enhanced emission}
\label{SI_sec:pol_response}
\add{The design of the nanoparticle-SAM-electrode is adapted from previous work \cite{chen_continuous-wave_2021} where the numercial simulation of the plasmonic modes was performed. (In Ref.~\cite{chen_continuous-wave_2021} the 2~\textmu m groove acted as an antenna at mid-IR wavelengths, which is not needed here). The simulations predict at least two distinct radiative modes at visible wavelengths that arise from the hybridization of the individual gap modes formed by each nanoparticle-on-mirror (NPoM) cavity between the nanoparticle and each side of the electrode \cite{baumberg_extreme_2019}.
Due to the added complexity in the electrically-contacted device studied here that often involved multiple nanoparticles in different positions within the nanogroove, we did not repeat the simulation of the plasmonic modes for the individual devices and refer the reader to Ref.~\cite{chen_continuous-wave_2021} for a general overview of the plasmonic response.} 

\begin{figure}[H]
		\centering
		\includegraphics[height=0.4\textwidth]{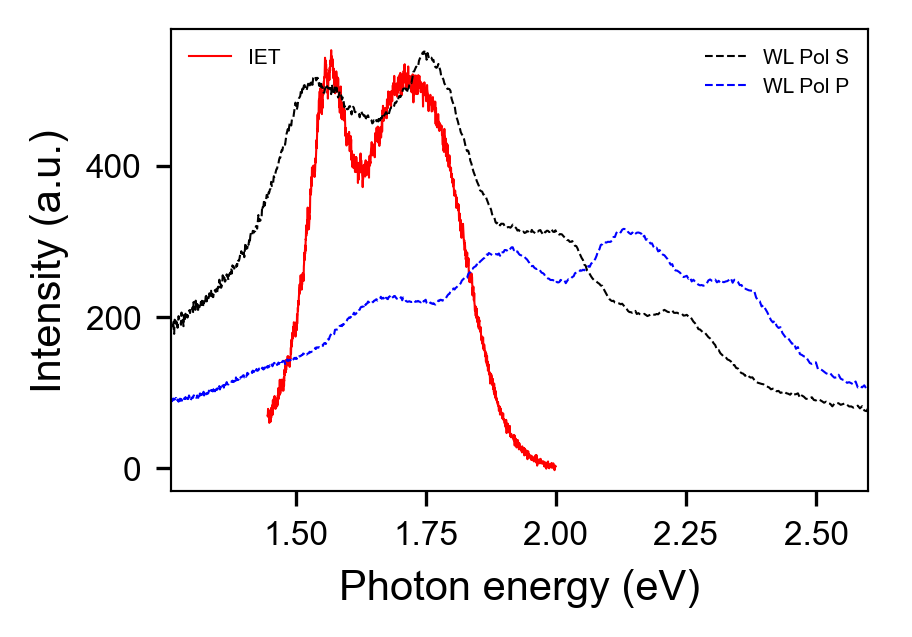}
		\caption{Light emission spectra at 1.9~V (red) and dark-field scattering spectra of a PMJ with S-(dotted black) and P-Polarized (dotted blue) excitation.}
  \label{figSI:PolDF}
\end{figure}

The electrical excitation of localized plasmons is confirmed from the dark-field scattering and the polarization response. We first obtain the dark-field scattering spectrum from the junction for P and S polarization of the incoming white light. The dark-field scattering spectroscopy is obtained using illumination from a tungsten lamp at about 15$^{\circ}$ incidence angle with respect to the sample plane as illustrated in Fig.~\ref{figSI:setup}. The spectrum is collected using a multimode fiber and measured with a QE pro-Ocean Insight spectrometer. The S-polarized excitation results in two dominant scattering peaks that are suppressedd under P polarization. The same peaks are excited by the IET process with an external bias voltage, indicating that they are of plasmonic origin. The dark-field scattering spectrum and IET emission of a PMJ at 1.9~V is shown in Fig.~\ref{figSI:PolDF} for mixed polarization. Some additional features in the dark-field spectrum are not found in the IET spectra because the dark-field scattering signal comes from the entire junction in the collection area including the nanocavities formed by multiple nanoparticles in the electrode gap. IET on the other hand couples to the plasmonic response of the conducting nanoparticle bridge only.

\begin{figure}[H]
		\centering
		\includegraphics[height=0.4\textwidth]{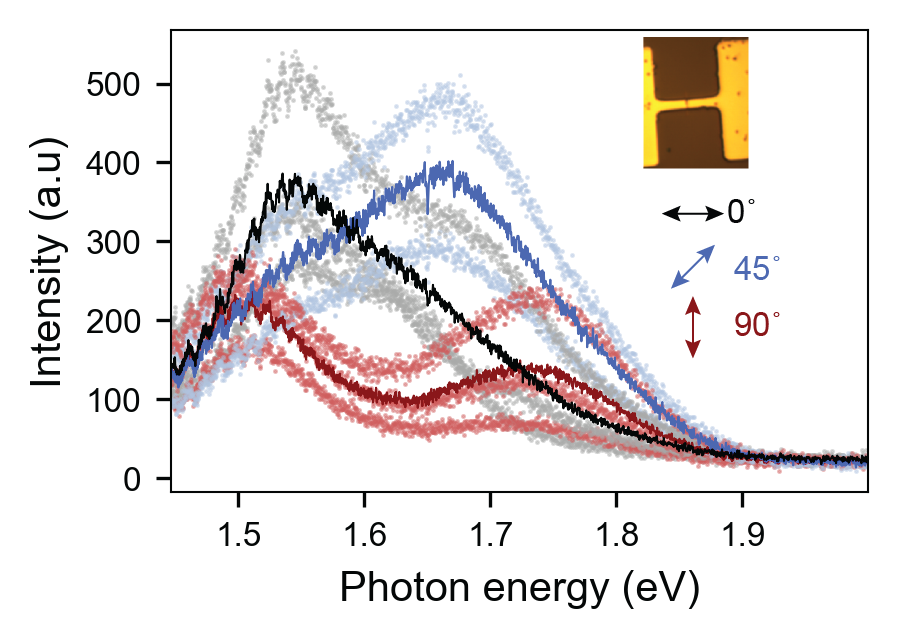}
		\caption{Light emission spectra at 1.8~V collected for various orientations of a polarizer in front of the spectrometer ($0^{\circ}$-black; $45^{\circ}$-blue; $90^{\circ}$-red represent the angle of the polarizer with respect to the electrode length), showing spectral reshaping of emission through the plasmonic cavity. Dark lines are the averages of individual exposures shown in faded lines. Inset - bright field image of the PMJ showing the orientation of the electrode. In all experiments of the main text, no polarizer is used in the detection path.}
  \label{figSI:Pol}
\end{figure}

To evidence the role of plasmonic resonances in the out-coupling of light emission, a linear polarizer is placed in front of the spectrometer, and the light emission spectrum is collected along different electric field orientations (Fig.~\ref{figSI:Pol}). The variation of the spectrum with the polarization angle is consistent with the enhancement and reshaping of emission by the plasmonic cavity. 

\begin{figure}[H]
		\centering
		\includegraphics[width =0.9\textwidth]{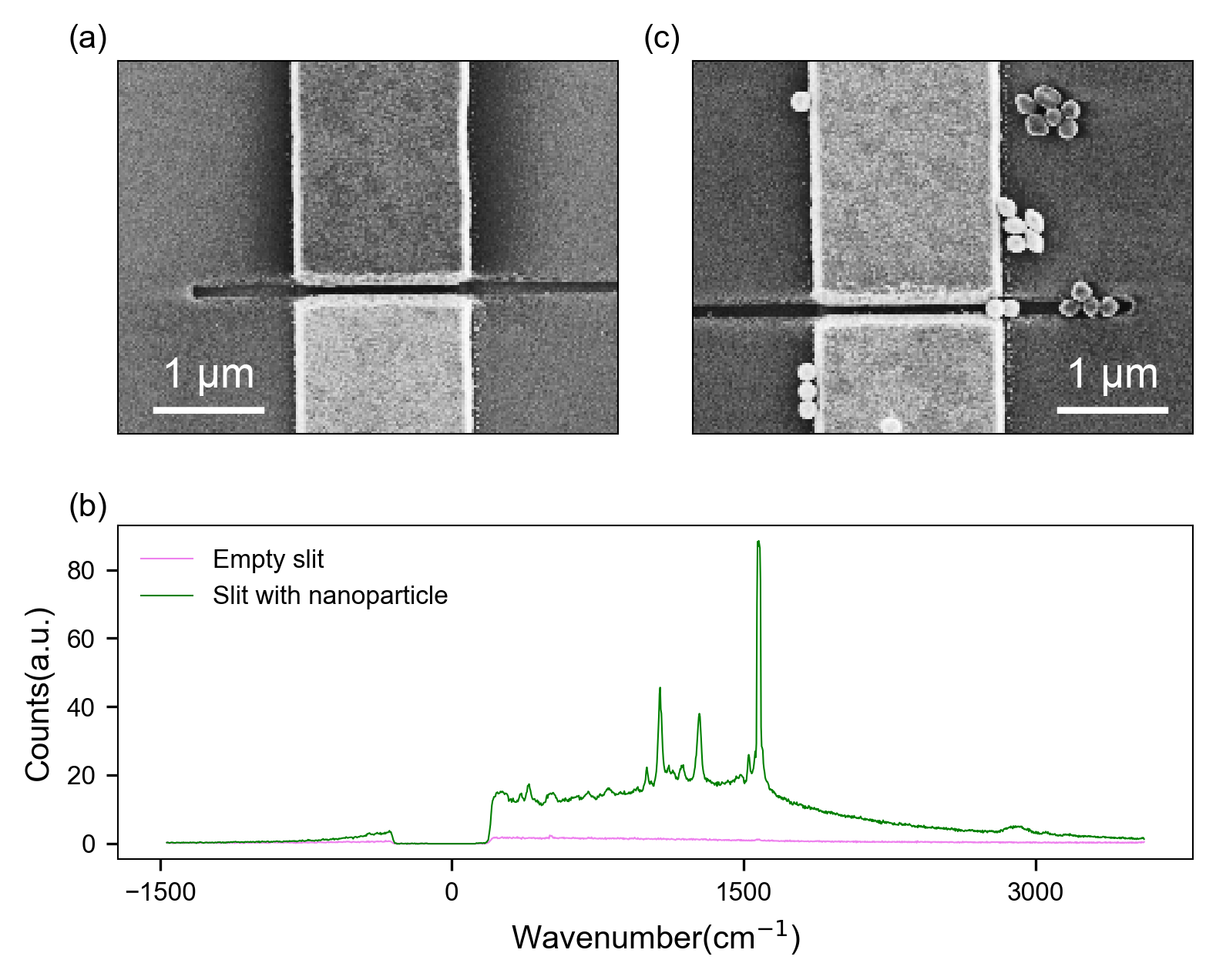}
		\caption{SEM image a PMJ with (a) electrode alone and (b) electrode with nanoparticle in the gap. (c) Their corresponding SERS spectra.}
  \label{figSI:SERS nanoparticle slit}
\end{figure}

The presence of an efficient plasmonic cavity is also evident from the observation of surface enhanced Raman scattering by the molecular monolayer. The plasmonic resonances at visible and NIR wavelengths \cite{chen_continuous-wave_2021} enable us to detect the vibrational signal from the molecules. In Fig~\ref{figSI:SERS nanoparticle slit}, we show the SERS signal of a PMJ with BPDT molecules. When no nanoparticle is present, the Raman signal is very weak. The presence of nanoparticles increases the electromagnetic field enhancement and also enhances emission from the molecular vibration, hence enhancing the SERS signal.

\section{Thermally assisted emission}
\label{secSI:Thermal emission}
\begin{figure}[H]
		\centering
		\includegraphics[height=0.4\textwidth]{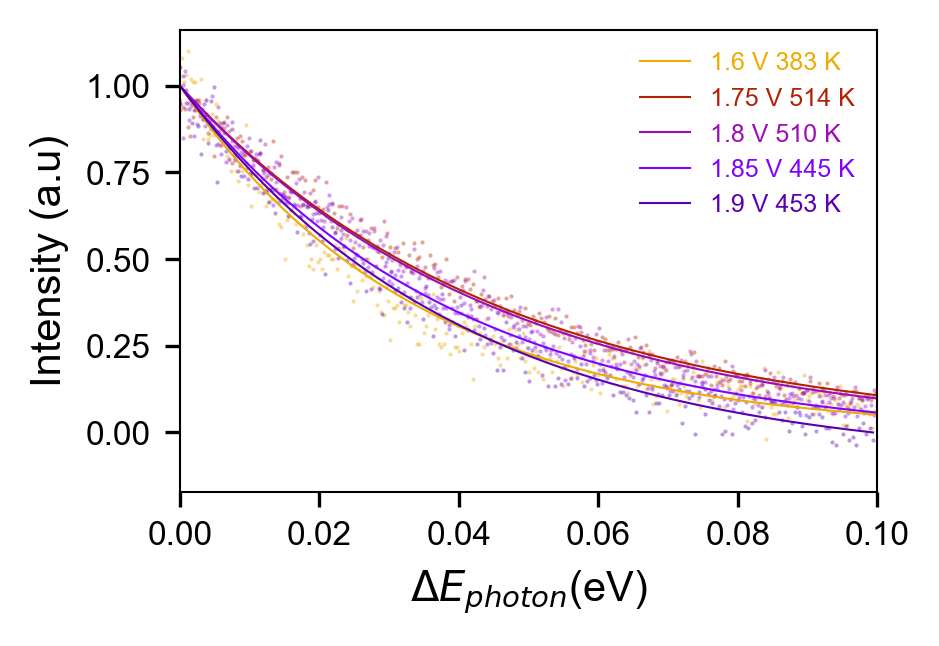}
		\caption{Thermal fit of the light emission spectra in the overbias region for various DC bias voltages (exposure time - 20~s). The energy values in the $x$-axis are computed with respect to the maximum photon energy corresponding to the bias voltage.}
  \label{figSI:overbias}
\end{figure}

The overbias contribution to the light emission spectra is attributed to the emission from a globally hot system at equilibrium. While it is possible for the electron temperature to be significantly higher than the lattice temperature, the observed overbias tails do not require such an assumption to be explained in our experiment. Fig.~\ref{figSI:overbias} shows the temperature estimated from the tail of the overbias emission with Boltzmann fit,
\begin{equation}
   I_{thermal}(\lambda) =  \frac{A. hc}{{\lambda.(exp(\frac{hc}{\lambda k_BT})-1})} + B
\end{equation}
where $\lambda$ is the wavelength, $h$ is Planck's constant, $c$ is the speed of light, $k_B$ is Boltzmann constant, $T$ is temperature, $A$ and $B$ are constants. 
Even without factoring out the plasmonic response (which is difficult to do properly due its fluctuating and voltage-dependent contribution) the thermal fit reasonably estimates slightly elevated temperatures of the PMJ compared to room temperature. 

\section{Stability of the devices}
\label{secSI:Stability of the devices}
\begin{figure}[H]
		\centering
		\includegraphics[width=0.8\textwidth]{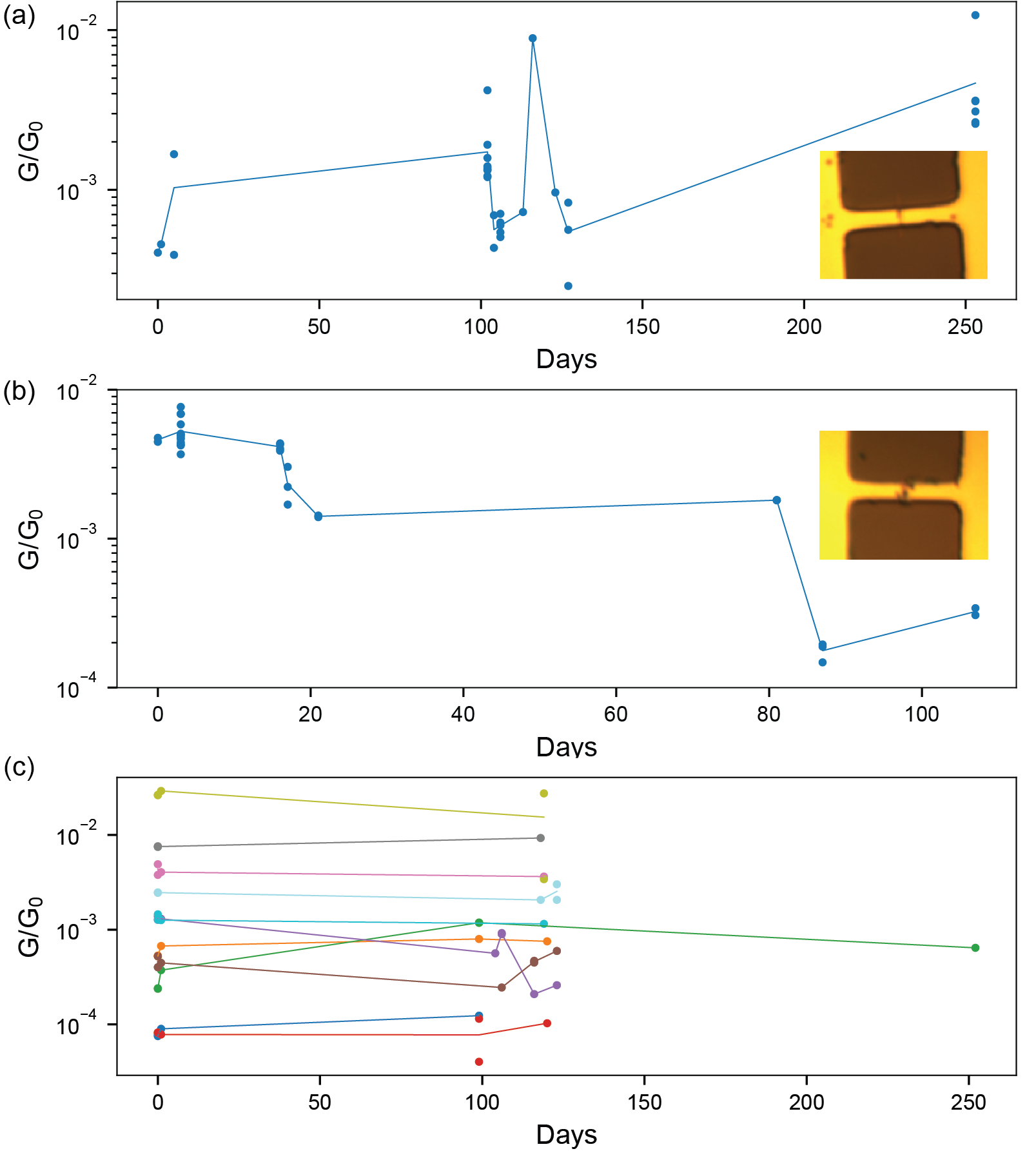}
		\caption{Conductance of two repeatedly measured PMJs (a)-(b) and several other PMJs (c) over several days (DC bias - 5~mV). Each color in (c) corresponds to an individual PMJ. The devices show excellent long-term survival.}
  \label{figSI:stability}
\end{figure}

The PMJ devices are working for several days and their conductances monitored over time in between measurements are shown in Fig.~\ref{figSI:stability}. The conductance state changes over time mostly \textit{during measurements} as shown for two devices in Fig.~\ref{figSI:stability}a-b. These two PMJs, for example, remained functional for more than 100 days.

\begin{figure}[H]
		\centering
		\includegraphics[width=0.8\textwidth]{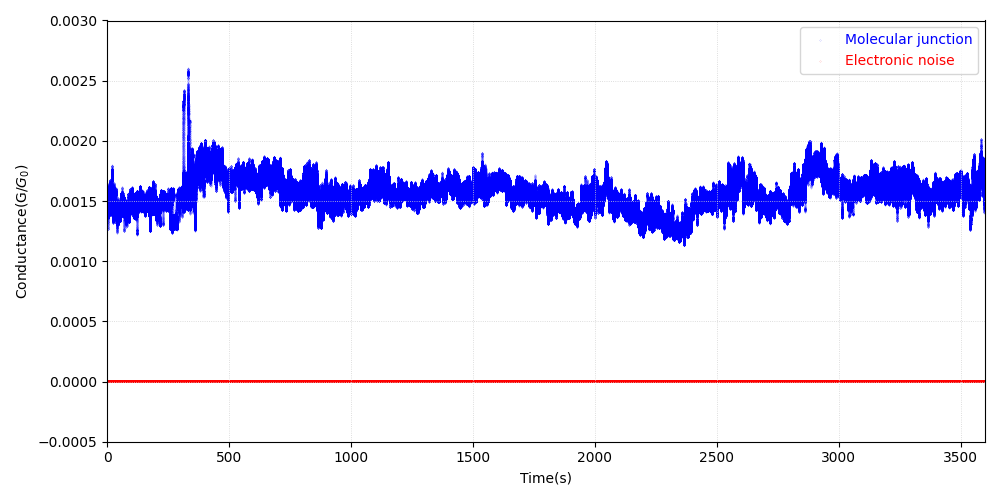}
		\caption{Stability of a PMJ for 1 hour (DC bias - 5~mV) in blue. Red line shows the open circuit noise from the electronic measurement unit.}
  \label{figSI:timetrace}
\end{figure}

A time trace of continuous conductance measurement for 1 hour is shown in Fig.~\ref{figSI:timetrace}

\section{Electrically induced fluctuations}
\label{secSI:voltage induced fluatuations}
\begin{figure}[H]
		\centering
		\includegraphics[width=\textwidth]{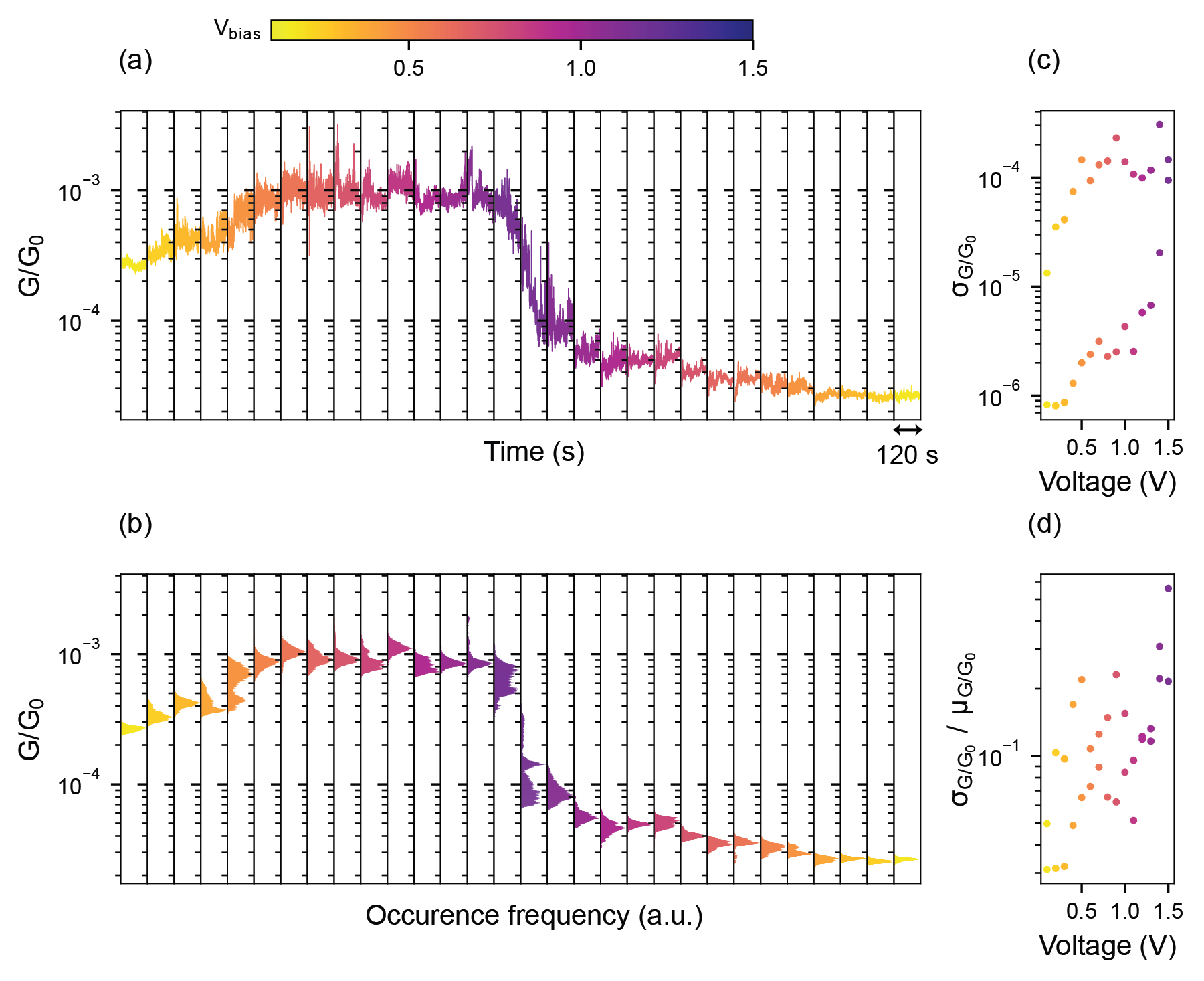}
		\caption{(a) Conductance of a BPDT junction measured for 120~s consecutively at each voltage from 0.1~V to 1.5~V and back to 0.1~V in steps of 100~mV. (b) Corresponding histograms of the conductance data. (c) Standard deviation ($\sigma$) of the conductance of each measurement in (a) plotted against the applied DC voltage. (d) The standard deviation of the conductance divided by its mean value ($\mu$) plotted against the voltage.}
  \label{figSI:bias_dependent_pico}
\end{figure}

The random fluctuations in the conductance are found to intensify with an increase in the applied bias voltages, as illustrated in Fig.~\ref{figSI:bias_dependent_pico}. The standard deviation of the conductance from each measurement is monotonously increasing with the increase in the voltage both in the forward and reverse sweep of voltages (Fig.~\ref{figSI:bias_dependent_pico}c). The device is prone to significant changes in its conductance state at high voltages as observed in the transition of conductance at 1.5~V. This indicates that fluctuations are strongly current- or voltage-driven, possibly involving a non-thermal mechanism given the very good heat dissipation provided by the large gold electrodes. 

Yet, we remark that once the standard deviation of the conductance is further normalized by its mean value (Fig.~\ref{figSI:bias_dependent_pico}d), the hysteresis from Fig.~\ref{figSI:bias_dependent_pico}c disappears. Two lines of explanation are proposed. First, it could be that more fluctuations correlate with a higher number of conducting channels. Second, it may indicate that more fluctuations correlate instead with the increase in the dissipated power through local Joule heating. Distinguishing the thermal and non-thermal contributions to the conductance fluctuations therefore requires further investigation of these effects with spectroscopic thermometry techniques or direct temperature scanning experiments.

\section{Details of the fit}
\label{secSI:Details of the fit}
To determine the photon emission as a function of conductance for a fluctuating PMJ, we model the intensity of light emission by the equation
\begin{equation}
    I_{PMJ}(t) = A\:.i(t)\:.f(V(t))
    \label{SIeq:photon-fit}
\end{equation}
where $A$ is a scaling factor to account for the overall efficiency of IET and photon detection, $i(t)$ is the current, and $f(V(t))$ is an experimentally determined function that accounts for the voltage-dependent photo-detection efficiency. 

As most of the voltage drop happens across one of the two junctions in series, the light emission is expected to originate from this dominating one. Hence, the conductance $G_1$ of the non-emitting junction is assumed to be much greater than $G_2$, the light emitting junction. Consequently, small fluctuations in $G_1$ have insignificant impacts on overall conductance and emission intensity. If $i(t)$ is the current flowing across this junction when the conductance $G_2$ is allowed to vary with time, then 
\begin{equation}
    V_{mid} - V_R = \frac{i(t)}{G_2(t)} 
    \label{eq:Vmid-G}
\end{equation}
\begin{equation}
    i(t) = V_{bias}.\frac{G_1. G_2(t)}{G_1+G_2(t)}
    \label{eq:i(t)}
\end{equation}
where $V_{bias} = V_L - V_R$. 

\begin{figure}[H]
		\centering
		\includegraphics[height=0.4\textwidth]{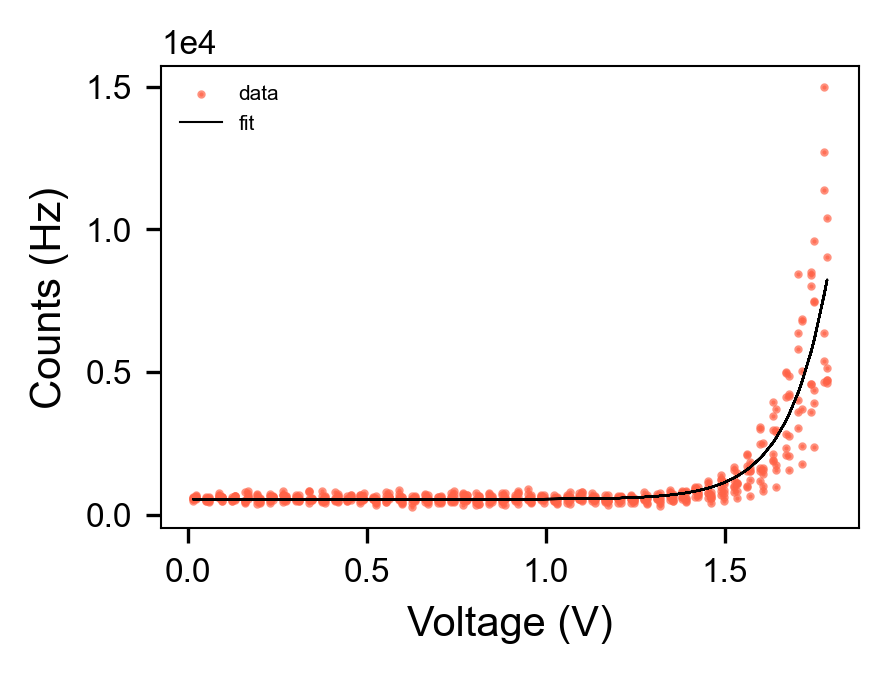}
		\caption{Experimental data of photon emission as a function of voltage (red dots) and the corresponding fit (black line).}
  \label{figSI:photonVsVoltage}
\end{figure}

\begin{figure}[H]
		\centering
		\includegraphics[width=\textwidth]{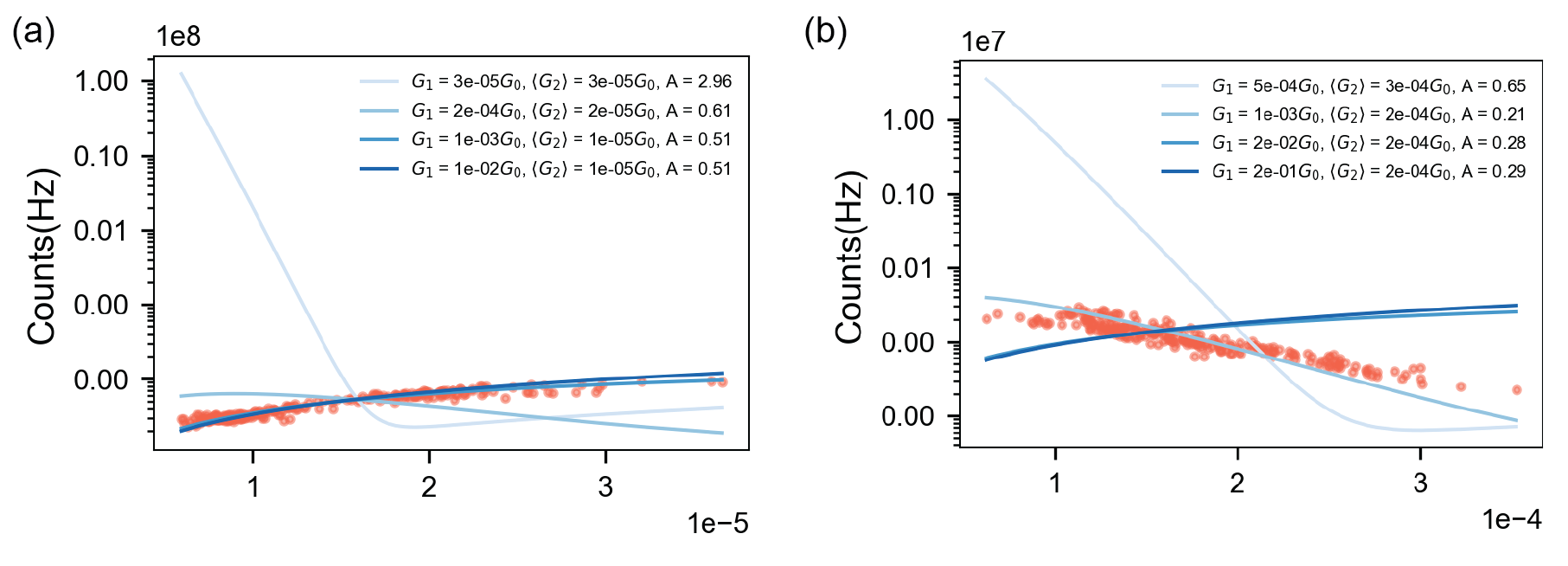}
		\caption{Multiple fits of experimental data discussed in the main text with different values for $G_1$, $G_2(t)$, and $A$ for (a) positively correlated and (b) inversely correlated regimes of conductance.}
  \label{figSI:model-fit}
\end{figure}

From the measured current values under constant DC bias, we obtain the values of the fit parameters $G_1$ and $G_2(t)$ in Equation.~\ref{eq:i(t)}. These parameters are used to derive the values of $V_{mid} - V_R$ for each value of the conductance, which in turn is used to estimate the values of $V_{L} - V_R$ as
\begin{equation}
    V_L - V_R = \frac{G_1+\langle G_2(t) \rangle}{G_1}.(V_{mid} - V_R)
\end{equation}
The voltage fluctuation $V(t) = V_L - V_R$ is then known. To estimate the photon counts from the voltage, we need to determine the function $f(V(t))$ in equation \ref{SIeq:photon-fit}. To obtain this, we measure the photon emission as a function of applied bias voltage (Fig.~\ref{figSI:photonVsVoltage}) and fit the curve with an equation,
\[
f(V) = \begin{cases}
C + B & \text{if } 0 \le V \le V_0 \\
C \cdot e^{k \cdot (V - V_0)} + B & \text{if } V > V_0
\end{cases}
\]
where $C$, $B$, $k$ and $V_0$ are fit parameters for the piecewise exponential function.
Finally, from the values of $f(V(t))$ and $i(t)$, the emission from PMJ can be fit with the equation \ref{SIeq:photon-fit} with a factor $A$. The fit obtained for different sets of the fitting parameters is shown in Fig.~\ref{figSI:model-fit}.

\section{Additional examples}
\label{SI_Sec:Additional examples}
\subsection{Light emission from PMJ with varied number of nanoparticles}
\begin{figure}[H]
		\centering
		\includegraphics[width=\textwidth]{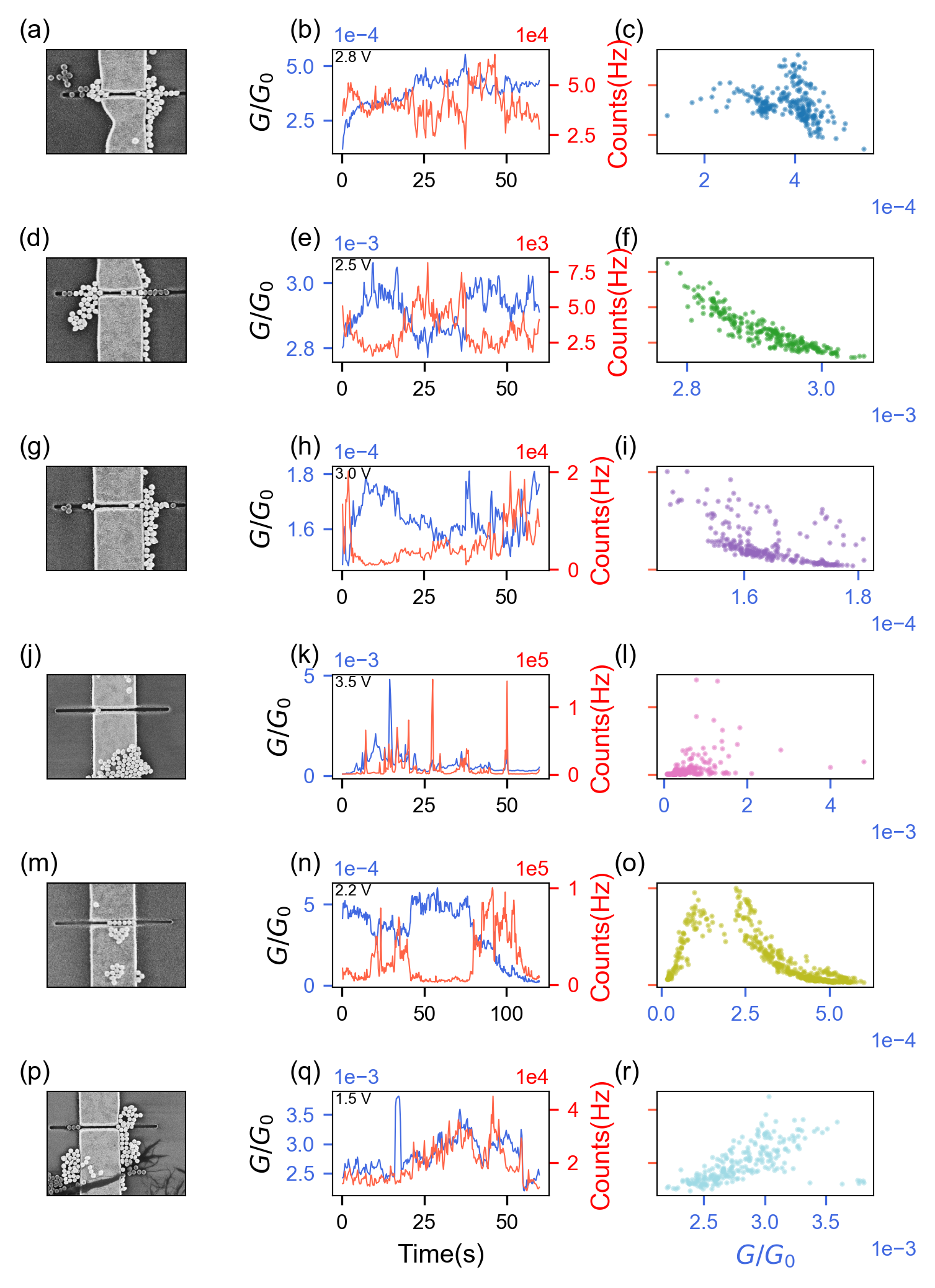}
		\caption{Correlatated fluctuations in conductance and light emission obtained from different PMJs with varied numbers of nanoparticles in the electrode gap along with their SEM image} 
  \label{figSI:SEM more data}
\end{figure}

\begin{figure}[H]
		\centering
		\includegraphics[width=\textwidth]{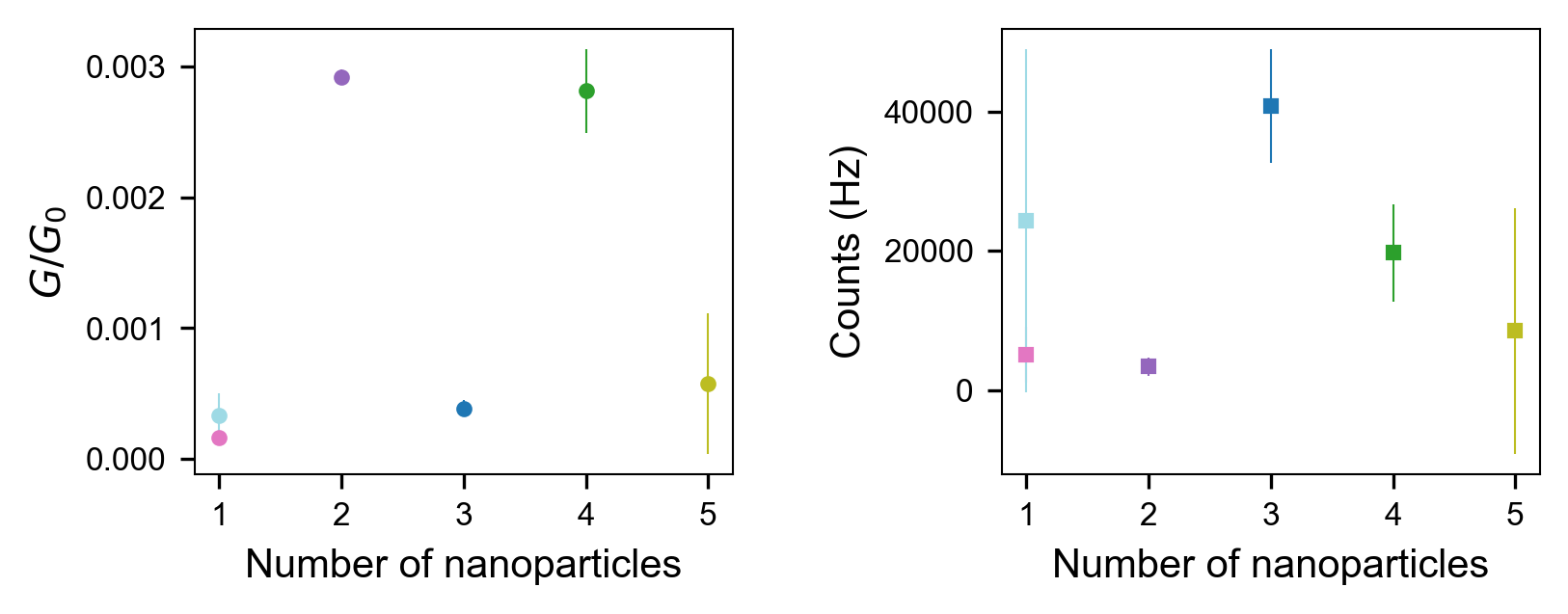}
		\caption{(a) Conductance and (b) Photon counts represented with errorbar for the devices in Fig.~\ref{figSI:SEM more data}, with the color of each data point corresponds to the color of the correlation scatter plot.} 
  \label{figSI:SEM more data summary}
\end{figure}

Correlatated fluctuations in conductance and light emission obtained from different PMJs with BPDT spacer and varied number of nanoparticles in the electrode gap are shown in Fig.~\ref{figSI:SEM more data}. 
\label{Sec: SEM summary}
Note that the measurements are performed at different D.C. voltages for each device to collect good signal from light emission. Each device could have a slightly varied threshold for light emission as discussed in Sec.~\ref{secSI:Light from IET}. The summary of fluctuations in conductance and light emission from these devices is shown in Fig.~\ref{figSI:SEM more data summary}. The conductance and light emission do not show any absolute scaling with the number of nanoparticles. Thus we conclude that despite the presence of multiple nanoparticles, only very few nanoparticles make successful electrical contact. Each device is slightly different in terms of conductance and light emission.

\subsection{PMJ with BPDT spacer}
\begin{figure}[H]
		\centering
		\includegraphics[width=0.95\textwidth]{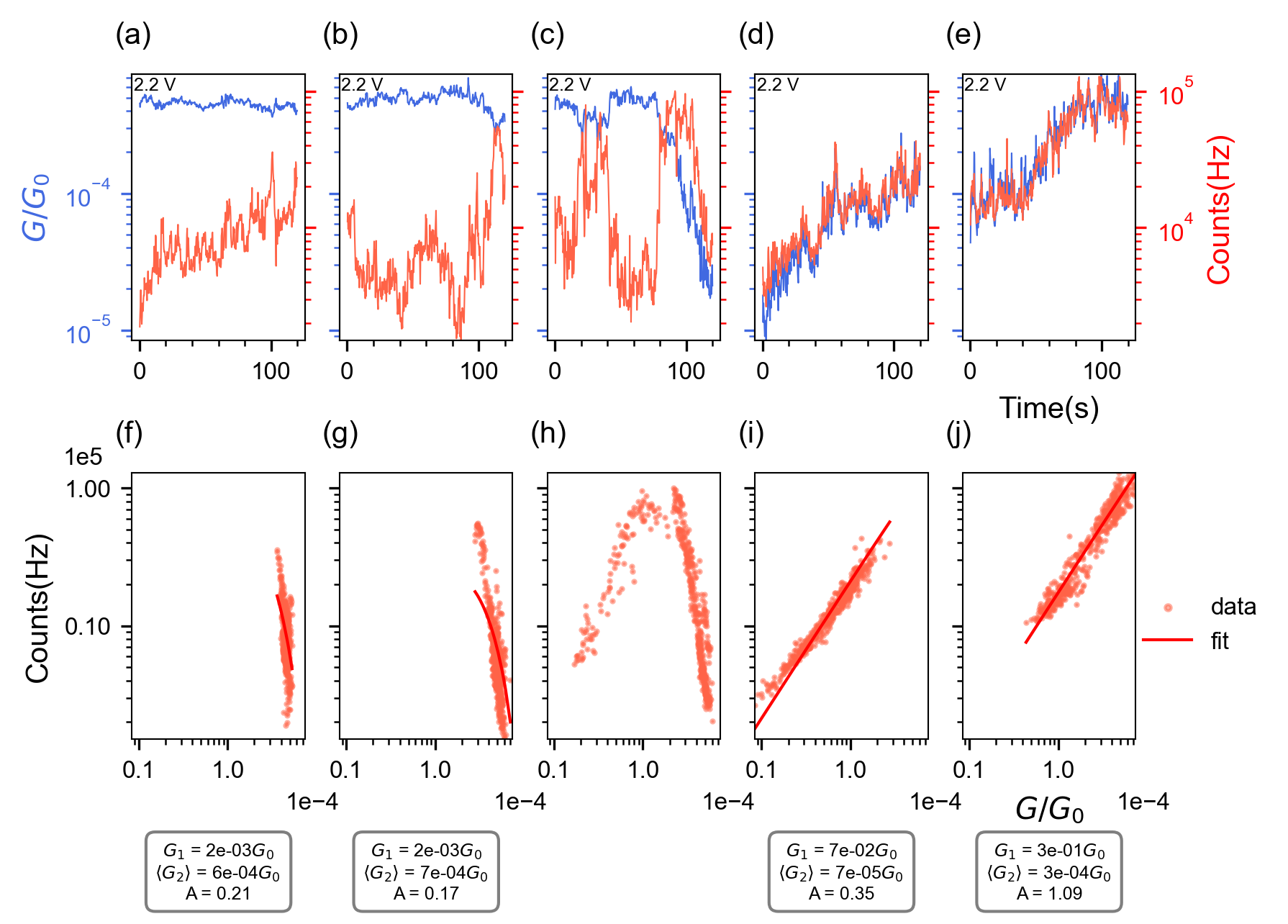}
		\caption{(a)-(e) Conductance (blue lines) and photon counts (red lines) simultaneously measured with SPCM for a particular PMJ with BPDT spacer. Both data sets are summed into 500~ms time bins. (f)-(j) corresponding correlation plots displaying the switching between positive and negative correlations. Solid line in each plot depicts the fit of the experimental data and the corresponding fit parameters are mentioned in the box below.} 
  \label{figSI:model-fit-bpdt}
\end{figure}

Another example of a PMJ showing the two regimes of conductance correlations is shown in Fig.~\ref{figSI:model-fit-bpdt} for a BPDT spacer. Noteworthy is the stability of conductance and emission intensity values observed between the end of a measurement and the start of the next one. This observation further suggests that the fluctuations are current- or voltage-driven, with little contribution from ambient thermal energy. 

\begin{figure}[H]
		\centering
		\includegraphics[width=0.9\textwidth]{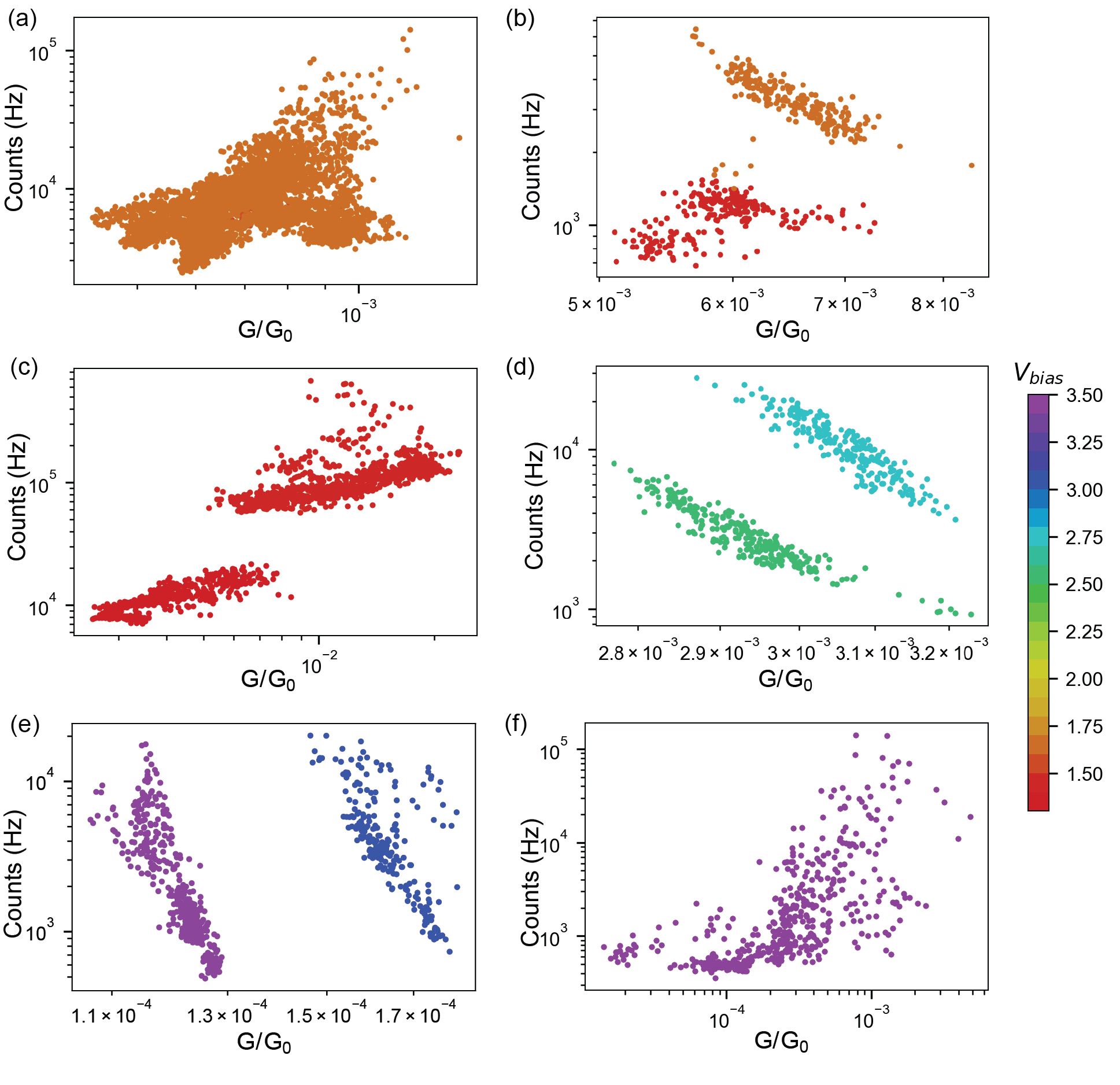}
		\caption{Correlation between photon count rate and conductance measured on different BPDT-spaced PMJs. Each plot from (a)-(f) represents an individual PMJ. The different colors indicate the corresponding applied DC bias.}
  \label{figSI:add-example-bpdt}
\end{figure}

More measurements performed at different DC bias voltages on a few other devices showing a monotonous (either positive or negative) conductance-emission correlation are shown as a summary in Fig.~\ref{figSI:add-example-bpdt}

\subsection{PMJ with citrate spacer}
\begin{figure}[H]
		\centering
		\includegraphics[width=0.95\textwidth]{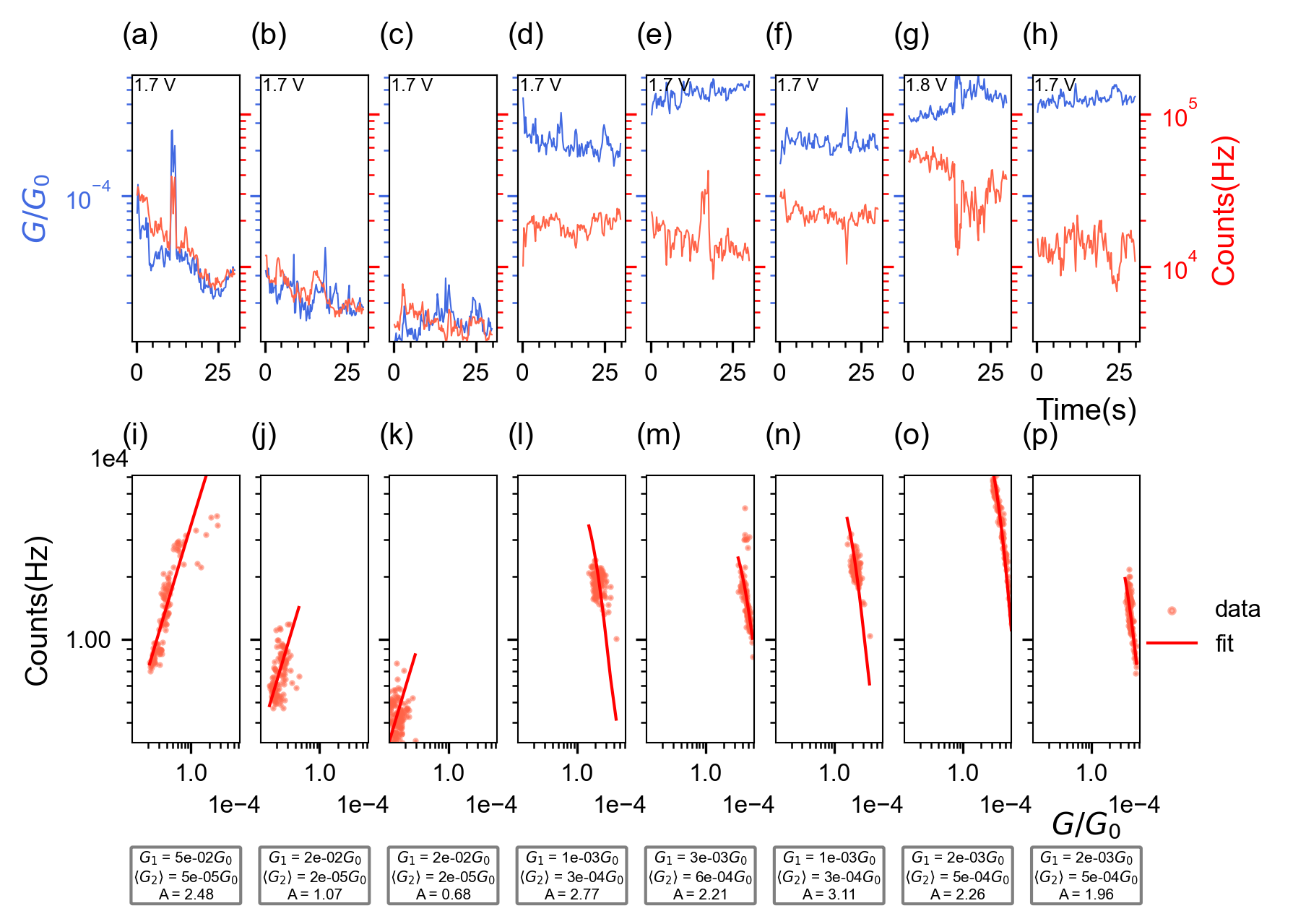}
		\caption{(a)-(h) Conductance (blue lines) and photon counts (red lines) simultaneously measured with SPCM for a particular PMJ with citrate spacer. Both data are summed into 500~ms time bins. (i)-(p) corresponding correlation plots displaying the switching between positive and negative correlations. Solid line in each plot depicts the fit of the experimental data and the corresponding fit parameters are mentioned in the box below.} 
  \label{figSI:model-fit-citrate}
\end{figure}

\begin{figure}[H]
		\centering
		\includegraphics[width=0.9\textwidth]{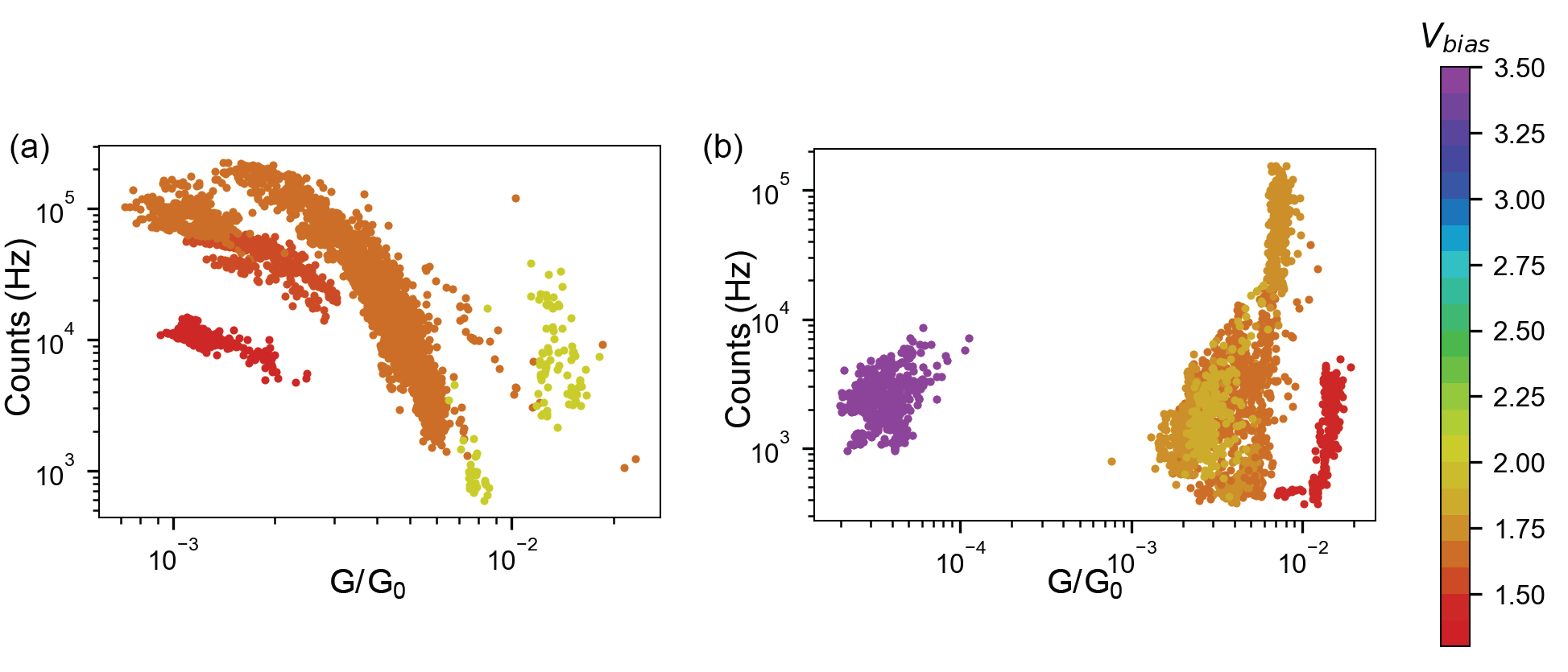}
		\caption{Correlation between photon count rate and conductance measured on a few other citrate-spaced PMJs. Each plot from (a)-(b) represents an individual PMJ. The different colors indicate the corresponding applied DC bias voltages.}
  \label{figSI:add-example-citrate}
\end{figure}

Some additional examples of PMJs with citrate spacer are shown in Fig.~\ref{figSI:model-fit-citrate} and Fig.~\ref{figSI:add-example-citrate}.

\section{Picocavity events in SERS and conductance}
\label{secSI: Picocavity events}
\begin{figure}[H]
		\centering
		\includegraphics[width=\textwidth]{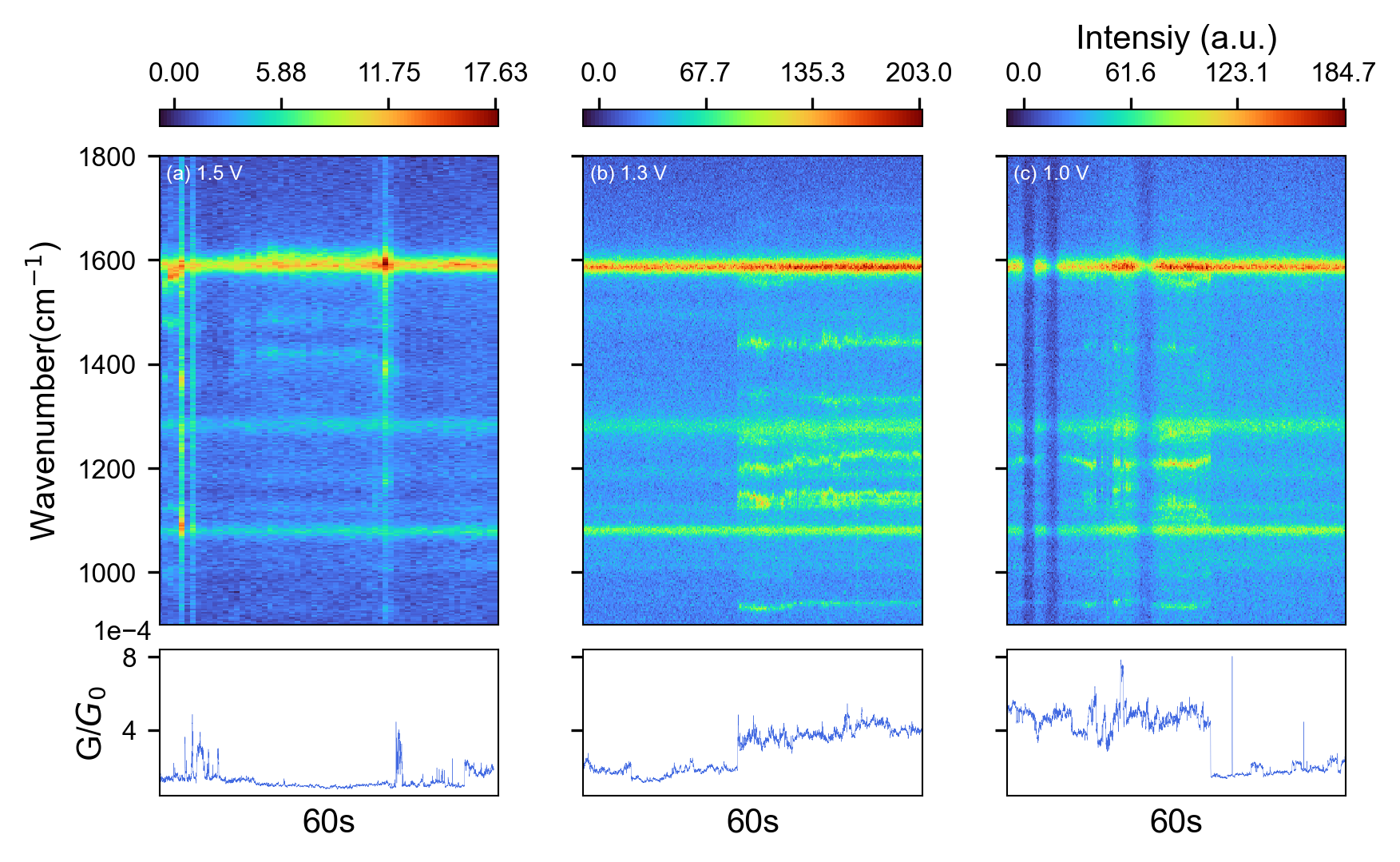}
		\caption{Time series of SERS spectrum along with the conductance showing picocavity events.}
  \label{figSI: picocavity timeseries}
\end{figure}

To show evidence of picocavity formation during intermittent blinking, we performed a combined conductance and SERS measurement on our PMJ. The picocavities in the nanogap create strong optical field gradients that modify the Raman selection rules and create additional vibrational peaks in the SERS spectra \cite{benz_single-molecule_2016}. In Fig.~\ref{figSI: picocavity timeseries}, we show a few picocavity events from PMJs that are correlated with the conductance jumps.

Such correlated events are very rare in SAM molecular junctions because we need to capture the conductance and SERS signal from the same molecule to observe the correlation. SERS in general is a collective signal obtained from all the molecules present in the nanogap, whereas conductance is obtained from a very few molecules connecting the electrodes. Thus the Raman signal is dominated by several other molecules that are not involved in the transport. Even in the presence of picocavities, it needs not occur from the same molecule that transmits the electrons. Such events are much less probable in PMJ as opposed to break junctions with atomic-sized tips \cite{ward_simultaneous_2008}. PMJs with single nanoparticle devices could be useful to perform such combined SERS and conductance measurements, where there will be better chances to observe direct evidence of picocavity events in transport.

\section{Outlook: Formation of single nanoparticle junction}
\label{secSI: DEP}
To trap a single nanoparticle in the nanogap, the dielectrophoresis (DEP) technique with feedback from optical imaging is used. DEP involves controlling a polarizable object with a non-uniform electric field \cite{pethig_review_2010}. When a particle is exposed to a non-uniform electric field, it is polarized and builds its own induced dipole moment. The DEP forces steer the particle towards the region of higher field intensity. The DEP force experienced by a spherical nanoparticle depends on several factors including the field intensity variation, the surrounding medium, and the size of the particle. Hence, DEP can be used to trap nanoparticles of varied concentration by applying an oscillating voltage of varied magnitude and frequency. The DEP forces can be expressed by the equation,
\begin{equation}
    <\textbf{F}_{DEP}(\omega)> = 2\pi \epsilon_m R^3 Re[f_{CM}(\omega)]\nabla|\textbf{E}_{rms}|^2
\end{equation}
where $\epsilon_m$ is the permittivity of the medium, $R$ is the radius of the nanoparticle, $f_{CM}(\omega)$ is the Clausius-Mosotti factor, and $E_{rms}$ is the rms value of the electric field. The Clausius-Mosotti factor describes the complex polarizability of the particle and is given by
\begin{equation}
    f_{CM}(\omega) = \frac{\epsilon_p - \epsilon_m -\frac{j}{\omega}(\sigma_p - \sigma_m)}{\epsilon_p + 2\epsilon_m -\frac{j}{\omega}(\sigma_p + 2\sigma_m)}
\end{equation}
where $\epsilon_p$ and $\sigma_p$ are the permittivity and the conductivity of the particle, while $\epsilon_m$ and $\sigma_p$ correspond to that of the medium, $\omega = 2\pi f$ is the angular frequency.

For the trapping to occur, the DEP forces have to be larger than the thermal motion described by,
\begin{equation}
    F_{th} = \frac{k_BT}{2R}
\end{equation}
where $k_B$ is the Boltzmann constant and $T$ is the temperature.

We ideally want to trap single nanoparticles in the gap. It has been shown that one could control the concentration of nanoparticles, voltage amplitude, and frequency of the oscillating field, and time of application of the voltage to manipulate the number of particles in the gap \cite{barsotti_assembly_2007, yoon_dielectrophoretic_2008, gierhart_frequency_2007, kumar_bridging_2009, cheon_assembly_2010}. 

\begin{figure}[H]
	\centering
	\includegraphics[width=\textwidth]{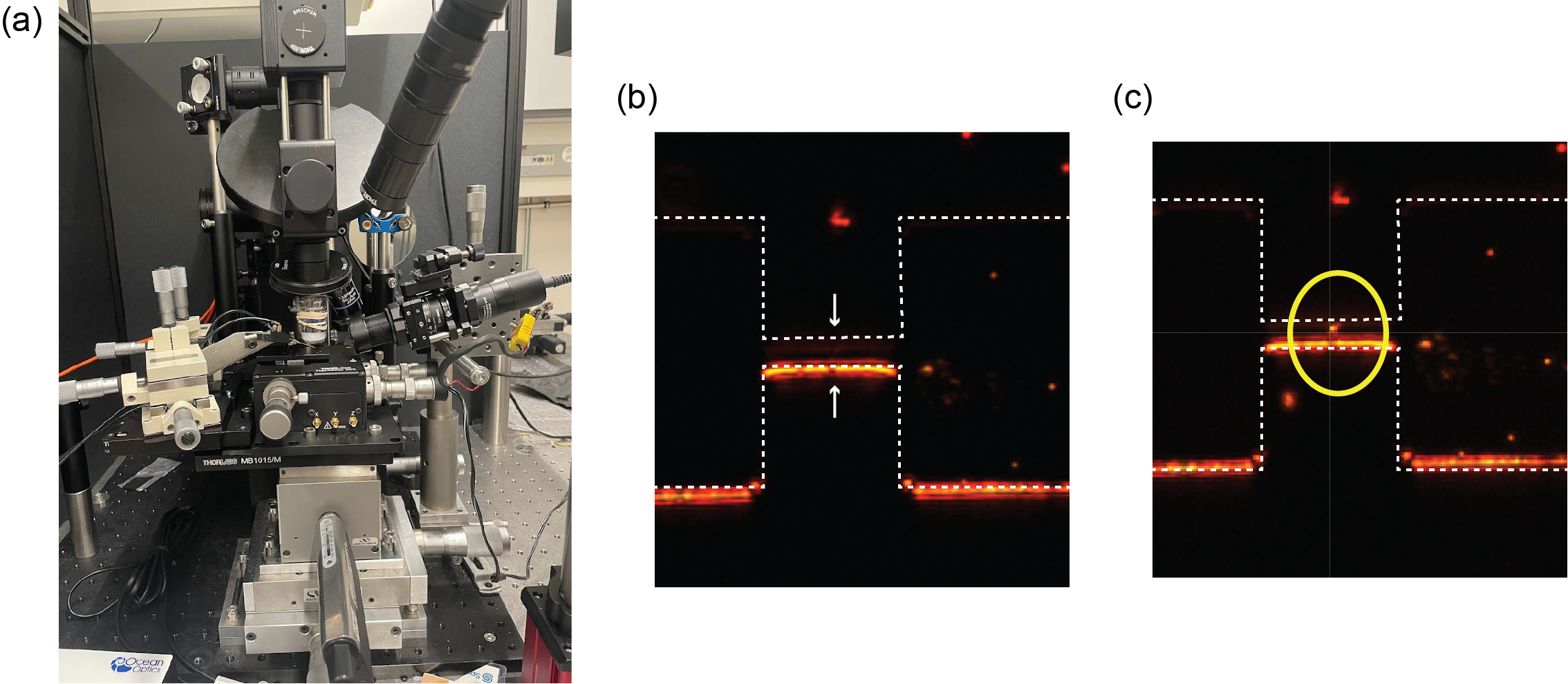}
	\caption{(a) Optical microscope with probes for DEP trapping. DF image of the electrode structures before (b) and after (c) trapping the nanoparticles by DEP. For better visibility, white dotted lines are overlayed on the DF image in (b,c) to mark the boundaries of the electrodes. White arrows point to the electrode gap. The yellow circle marks the location of the trapped particle.}
	\label{fig: dep single np}
\end{figure}

An optical microscope was modified to include probes that are attached to the sample holder (Fig.~\ref{fig: dep single np}a). The probes can contact the chip directly for the DEP experiment. A water immersion objective is used to observe the dark-field image throughout the trapping process. About 50~$\mu l$ of DI water is placed between the objective and the sample for imaging. 8~$\mu l$ of 10~OD nanoparticle solution in DI water is added to the solution. AC voltage between 1-3~V with a frequency of 500~kHz - 1~MHz is used to trap the nanoparticles. Different samples (normally with varied electrode widths and nanogap) require slightly different voltages and frequencies to trap the particles. The trapping can be monitored live from dark-field imaging. The DF image before and after trapping the nanoparticles is shown in Fig.~\ref{fig: dep single np}b,c.
The AC voltage is stopped once the particles are trapped, and the excess solution is blow-dried. Sometimes the particles are trapped with AC voltage but get released when the voltage is turned off. In such cases, a small DC voltage of 100~mV helps to stick the particles in the gap. With this approach, we could obtain devices with almost $100\%$ yield. Sometimes, if the image was not clear enough due to excess scattering from the electrodes, we might end up with more than one nanoparticle. Otherwise, a single nanoparticle junction could be achieved reproducibly. 
\begin{figure}[H]
	\centering
	\includegraphics[width=0.6\textwidth]{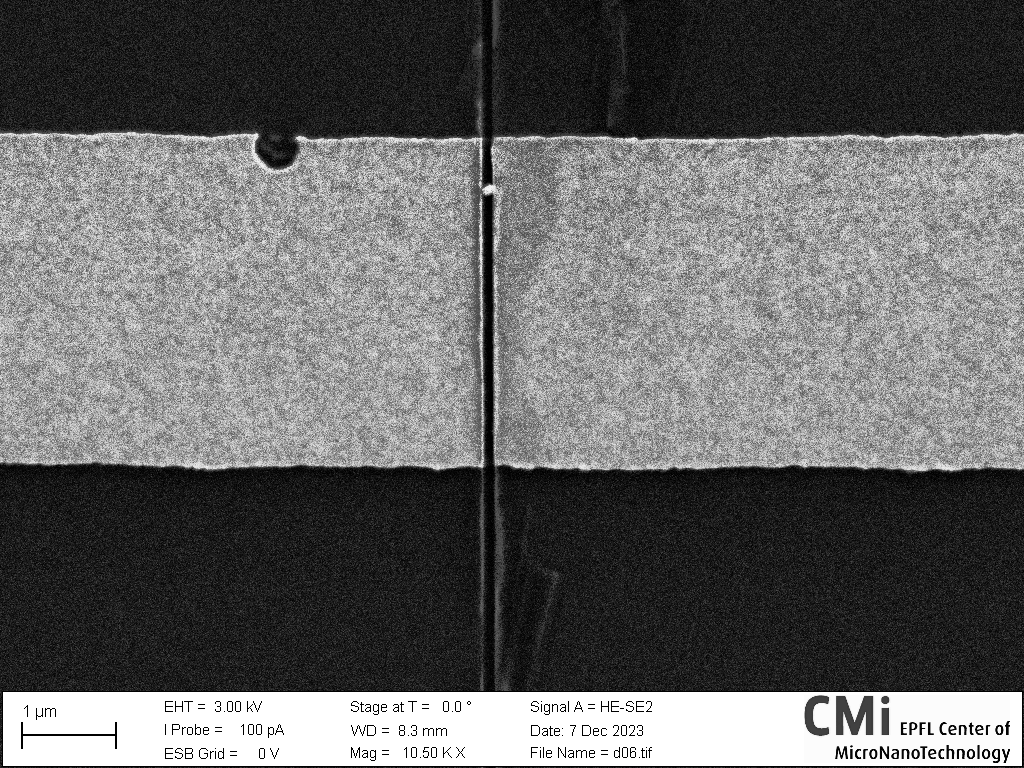}
	\caption{Single nanoparticle junction}
	\label{fig: single np}
\end{figure}
When the electrodes are functionalized with BPDT molecules, it increases the natural deposition rate of the molecules. This is not suitable for optical measurements as many particles stick in the vicinity of the electrode gap. To avoid this, one could try molecular functionalization and nanoparticle deposition without removing the PMMA layer (e-beam resist).

\textbf{Note:} These are preliminary results. The devices used for experiments discussed in this manuscript are from junctions with a few nanoparticles bridging the electrodes as described in Sec.~\ref{secSI: Formation of molecular spacer and nanoparticle bridge}.

\bibliography{references.bib}

\end{document}